\documentclass{aa}

\usepackage{graphicx}
\usepackage{txfonts}
\usepackage{color}

\begin{document}

\title{The effect of warm gas on the buckling instability in galactic bars}

\author{Ewa L. {\L}okas
}

\institute{Nicolaus Copernicus Astronomical Center, Polish Academy of Sciences,
Bartycka 18, 00-716 Warsaw, Poland\\
\email{lokas@camk.edu.pl}}


\abstract{
By using $N$-body and hydro simulations, we study the formation and evolution of bars in galaxies with significant gas
content focusing on the phenomenon of the buckling instability. The galaxies are initially composed of a spherical dark
matter halo and only stellar, or stellar and gaseous, disks with parameters that are similar to the Milky Way
and are evolved for 10 Gyr. We consider different values of the gas fraction $f =0-0.3$ and in order to
isolate the effect of the gas, we kept the fraction constant during the evolution by not allowing the gas to cool and
form stars. The stellar bars that form in simulations with higher gas fractions are weaker and shorter, and
they do not form at all for gas fractions that are higher than 0.3. The bar with a gas fraction of 0.1 forms sooner due
to initial perturbations in the gas, but despite the longer evolution, it does not become stronger than the one in the
collisionless case at the end of evolution. The bars in the gas component are weaker; they reach their maximum strength
around 4 Gyr and later decline to transform into spheroidal shapes. The distortion of the stellar bar during the
buckling instability is weaker for higher gas fractions and weakens the bar less significantly, but it has a similar
structure both in terms of radial profiles and in face-on projections. For $f=0.2,$ the first buckling lasts
significantly longer and the bar does not undergo the secondary buckling event, while for $f=0.3,$ the buckling does
not occur. Despite these differences, all bars develop boxy/peanut shapes in the stellar and gas component by
the end of the evolution, although their thickness is smaller for higher gas fractions.}


\maketitle

\section{Introduction}

A significant fraction of late-type galaxies in the Universe is barred \citep{Knapen2000, Aguerri2009, Nair2010,
Buta2015, Erwin2018}. This morphological feature has been studied over the last few decades both observationally and
theoretically, mainly via computer simulations. It has been established that bars can form via inherent instability of
galactic disks \citep{Hohl1971, Ostriker1973, Sellwood1981} when evolving in isolation and they grow by transferring
angular momentum from the disk to the halo \citep{Athanassoula2003}. Bars can also be induced in disks that are stable
in isolation if they are subject to strong enough tidal forces from neighboring structures \citep{Noguchi1996,
Lokas2014, Lokas2016, Gajda2017, Lokas2018}.

A common episode in the evolution of strong bars, independent of their origin, is the phenomenon of buckling
instability. The instability, which was first discussed by \citet{Combes1981}, \citet{Pfenniger1991}, and
\citet{Raha1991}, manifests itself by noticeable distortions of the bar out of the disk plane that modify the orbital
structure of the bar producing a pronounced boxy/peanut shape of the bar when seen edge-on, which is quite similar to
those observed \citep{Erwin2017}. The first rather short buckling event occurs in the inner part of the bar and can
also then extend itself to the outer region where it lasts much longer \citep{Martinez2006}.

The nature of the instability remains controversial; it has been ascribed to the so-called fire-hose instability and is
then supposed to be triggered by a low enough ratio of the vertical to horizontal velocity dispersion of the stars in
the bar \citep{Toomre1966, Merritt1991, Merritt1994}. Another possibility, first proposed by \citet{Combes1990} and
\citet{Pfenniger1991}, and more recently by \citet{Quillen2002}, is that the instability results from trapping x1 orbits
of the bar at vertical resonances. However, these resonances should mainly apply to banana-like orbits, which are known
to only contribute a little to the final boxy/peanut shape \citep{Portail2015, Abbott2017}.

In \citet{Lokas2019a}, by using simulations of tidally induced bars, we demonstrate that there is no direct relation
between the ratio of the vertical to horizontal velocity dispersion and the occurrence of buckling, which is in
opposition to the interpretation of buckling as the fire-hose instability. In \citet{Lokas2019b} we study the buckling
instability in a collisionless simulation of a Milky Way-like galaxy, looking in particular at the evolution of
the orbital structure of the bar due to buckling. It turns out that buckling seems to be indeed initiated by the
vertical instability of x1 orbits that cause the initial distortion of the bar out of the disk plane. This distortion
then evolves as kinematic bending waves changing the orbital structure into a combination of much more
complicated orbits; however, they all obey a very tight relation between the vertical and circular frequency.

The effect of the dissipative component of galaxies on the evolution of bars, and the buckling instability in
particular, has so far been studied mainly in the context of the discussion concerning the bar survival or longevity.
The controversy started with the results of \citet{Bournaud2002} and \citet{Bournaud2005} who claimed, based on their
simulations, that the bars in gas-rich disks are short-lived because of the formation of the central mass concentration
and the torques between the stellar bar and the gaseous arms. Later studies \citep{Debattista2006, Berentzen2007,
Villa2010, Athanassoula2013} disagreed with these conclusions and showed that bars can survive in galaxies with
significant gas content and their temporary weakening is due to the buckling instability. These studies, however,
demonstrated that bars are generally weaker in gas-rich galaxies and indicated that buckling instability is then less
pronounced.

In this paper, we revisit the issue of the buckling instability with the aim of isolating the effect of the presence of
a significant fraction of warm gas in the disk. In order to control the gas content, it seems reasonable to keep it
constant by not allowing it to cool and form stars. Otherwise, as demonstrated by simulations of
\citet{Athanassoula2013}, the gas is quickly depleted. They considered galaxies with gas fractions from a wide range
initially, but these fractions, due to star formation, turned out to be quite similar at the most interesting time when
the bar forms. While it may seem more realistic to include cooling, star formation, and feedback, these processes do
not take into account the continuous supply of gas from the hot halo, which has been demonstrated to significantly
affect the evolution of galaxies \citep{Roskar2008}. Given this, the assumption of a constant gas fraction may not be
as unrealistic as it seems and it may reflect, to some extent, the actual gas content history in real galaxies. Our
approach is thus more similar to the one of \citet{Berentzen2007} and \citet{Villa2010} who considered constant gas
fractions. However, in contrast to their studies, we used a higher resolution for the gas, a novel method to solve
hydrodynamics, and evolved the gas adiabatically instead of keeping it isothermal. We also applied newly developed
methods to quantify and visualize buckling instability.

The paper is organized in the following way. In Section 2 we present the simulations used in this study. Section 3
describes the formation and evolution of the bars and Section 4 gives a more detailed description of buckling in the
presence of a dissipative component. The discussion follows in Section 5.

\section{Simulations}

\begin{figure}
\centering
\includegraphics[width=7cm]{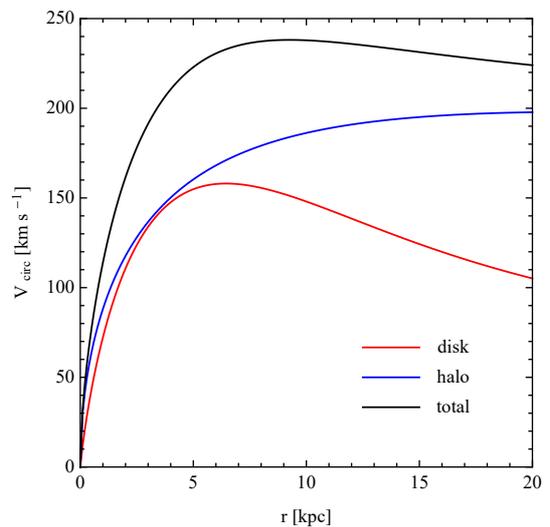}
\caption{Initial circular velocity curve of the simulated galaxies. The red, blue, and black curve corresponds to the
disk, halo, and both components combined, respectively.}
\label{circvel}
\end{figure}

For the purpose of this study, we ran a series of simulations of a galaxy similar to the Milky Way evolving in
isolation. The initial conditions for each simulation were the same, except for the gas fraction in the disk. All
galaxy models were initially composed of a spherical dark matter halo and a purely stellar exponential disk, in the
collisionless case, or a stellar and a gaseous disk, in the remaining cases. The dark matter halo had a
Navarro-Frenk-White \citep{Navarro1997} profile with a virial mass of $M_{\rm H} = 10^{12}$ M$_{\odot}$ and a
concentration of $c=25$. The total mass of the disk (whether purely stellar or composed of stars and gas) was $M_{\rm
D} = 4.5 \times 10^{10}$ M$_{\odot}$, its targeted scale-length was $R_{\rm D} = 3$ kpc (for both components), and the
thickness was $z_{\rm D} = 0.42$ kpc (for the stars). The central value of the radial velocity dispersion for the stars
was $\sigma_{R,0}=120$ km s$^{-1}$. The initial circular velocity curve of the simulated galaxies is shown in
Fig.~\ref{circvel}.

For the models containing gas, the mass of the stellar disk was reduced, the mass of the gaseous disk was increased,
and we considered different gas fractions $f$ of the disk, from 0 to 0.7 with a step of 0.1. All models had $10^6$ dark
matter particles in the halo (of mass $10^6$ M$_\odot$ each) and $10^6$ particles in the disk. The masses of the
stellar and gas particles were always the same and equal to $4.5 \times 10^4$ M$_\odot$. So, for example, the disk with
the gas fraction $f=0.1$ had $10^5$ gas particles and $9 \times 10^5$ stars.

The galaxies were initialized using the new version of the GalactICS code \citep{Deg2019} that extends the procedures
described in \citet{Widrow2005} and \citet{Widrow2008} to include gaseous disks. The code uses the potential method of
\citet{Wang2010} to create gaseous disks close to equilibrium. The collisional component was assumed to be an ideal,
adiabatic gas with a temperature of $10^4$ K. The gas disks that were created were thin and flared toward larger radii,
as expected for systems in equilibrium \citep{Benitez2018}. To quantify the flaring, we estimated the median value of
the vertical $|z|$ coordinates of the gas particles as a function of the cylindrical radius $R$. We find that the
medians increase in radius with values of 0.03, 0.06, and 0.1 kpc at $R = R_{\rm D}$, $2 R_{\rm D}$, and $3 R_{\rm D}$,
respectively, and they are very similar for all gas fractions. Instead, the analogous medians of $|z|$ for the stellar
disks are almost constant with radius and are equal to 0.22-0.24 kpc in the same range of radii.

\begin{figure}
\centering
\includegraphics[width=4.4cm]{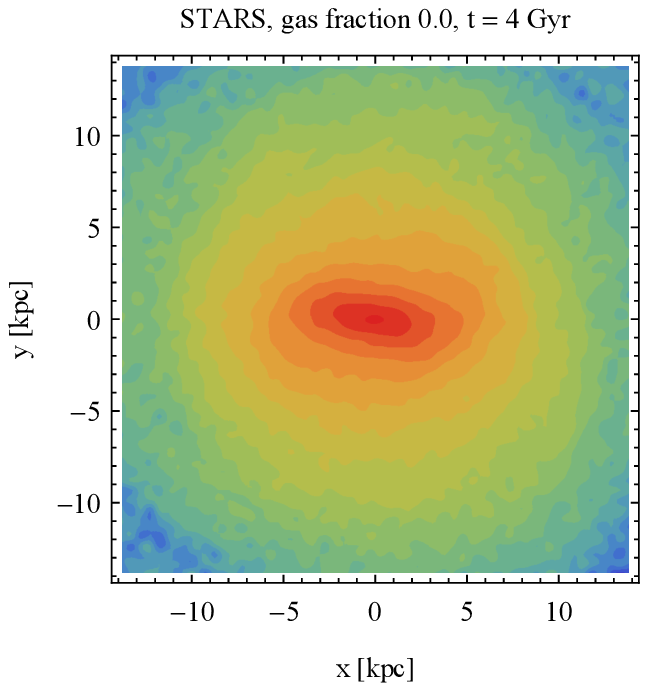}
\includegraphics[width=4.4cm]{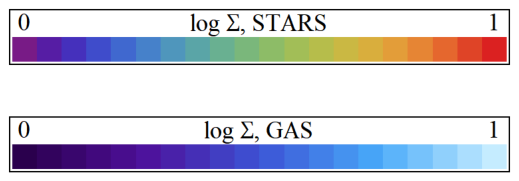} \\
\vspace{0.3cm}
\includegraphics[width=4.4cm]{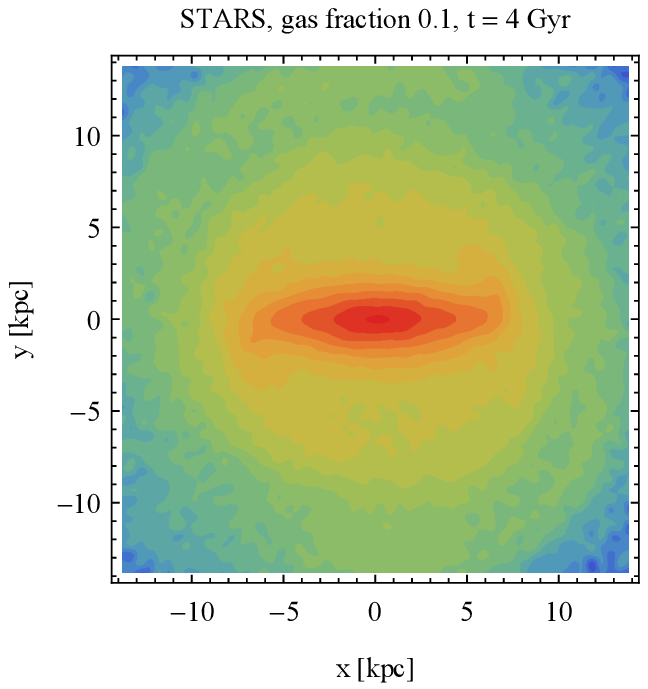}
\includegraphics[width=4.4cm]{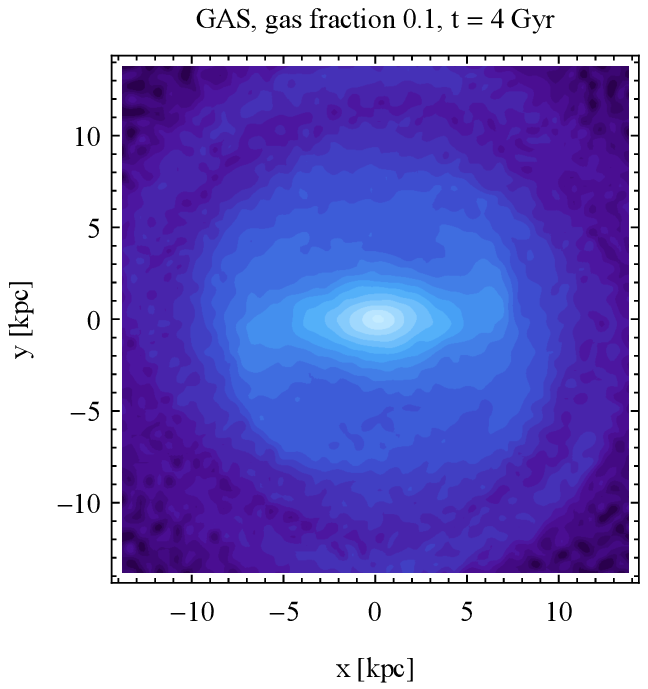} \\
\vspace{0.3cm}
\includegraphics[width=4.4cm]{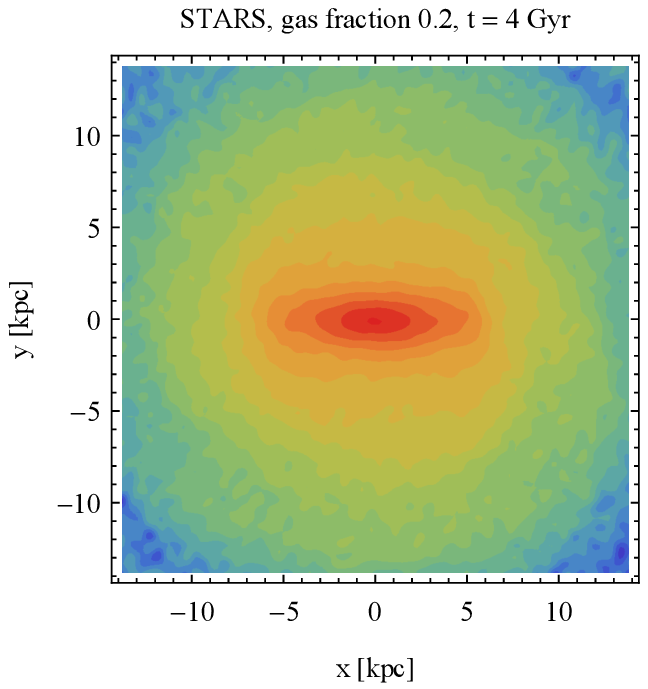}
\includegraphics[width=4.4cm]{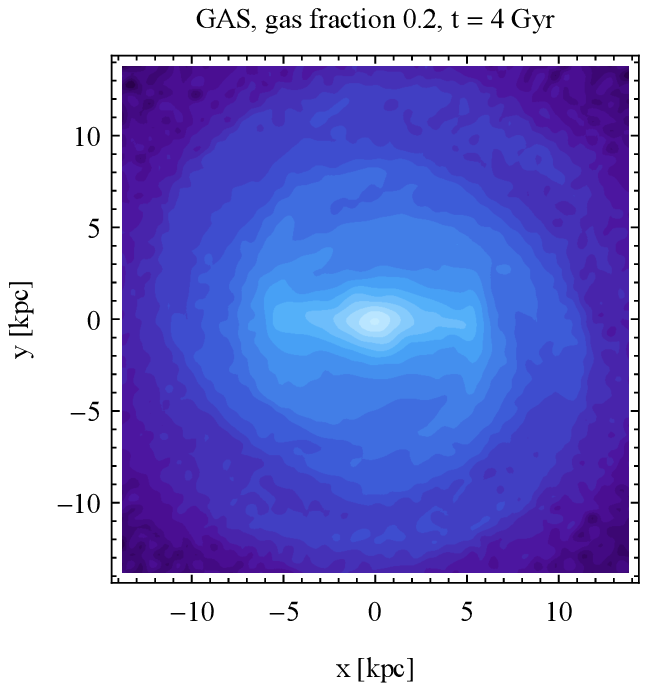} \\
\vspace{0.3cm}
\includegraphics[width=4.4cm]{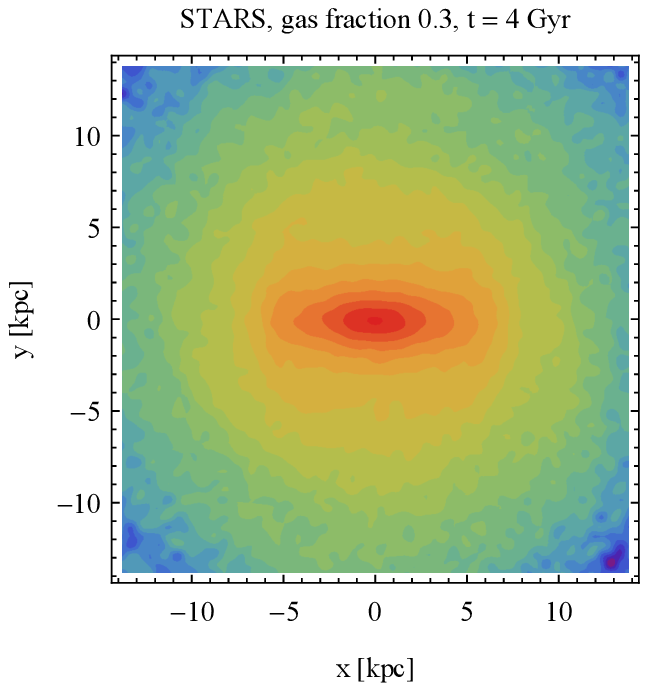}
\includegraphics[width=4.4cm]{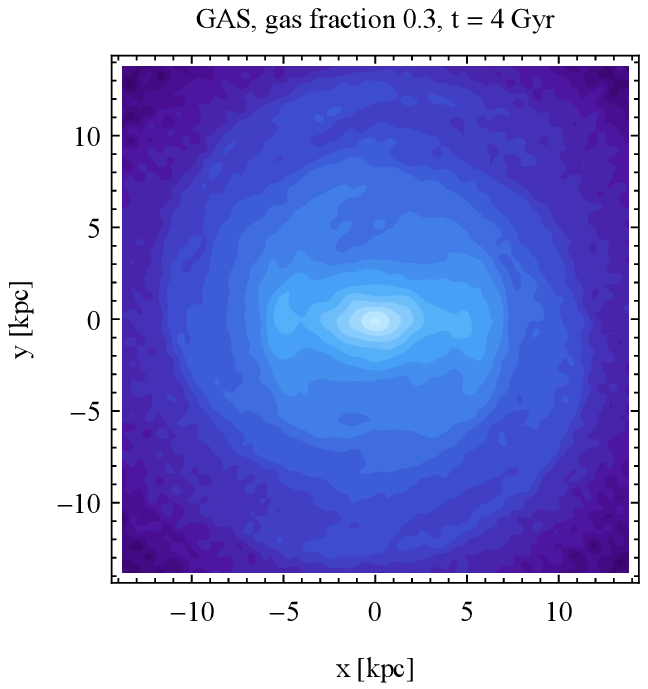}
\caption{Surface density distributions of stars and gas in the face-on view at 4 Gyr from the start of the simulations
when the gas bars are strongest. Rows show the images from simulations with increasing gas fraction. The left column is
for stars and the right one is for the gas. The surface densities were normalized to the central value in each case.}
\label{surdenxy4gyr}
\end{figure}

\begin{figure}
\centering
\includegraphics[width=4.4cm]{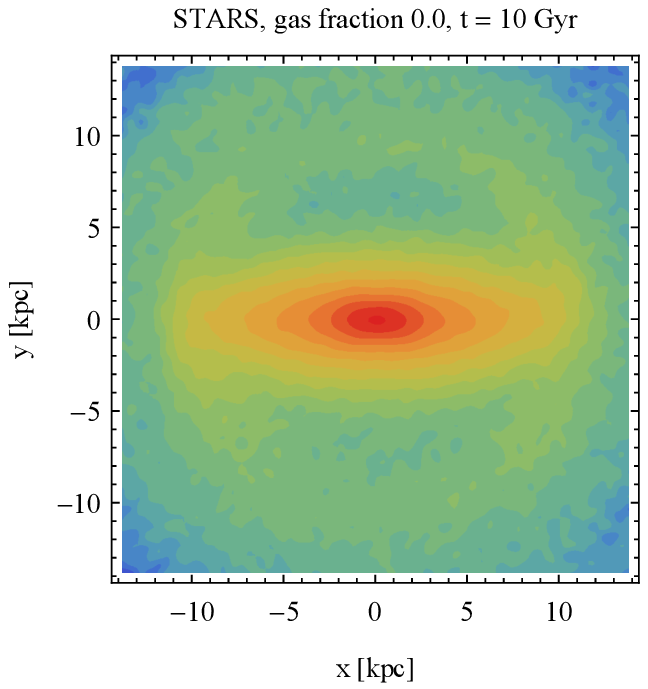}
\includegraphics[width=4.4cm]{legend19a.eps} \\
\vspace{0.3cm}
\includegraphics[width=4.4cm]{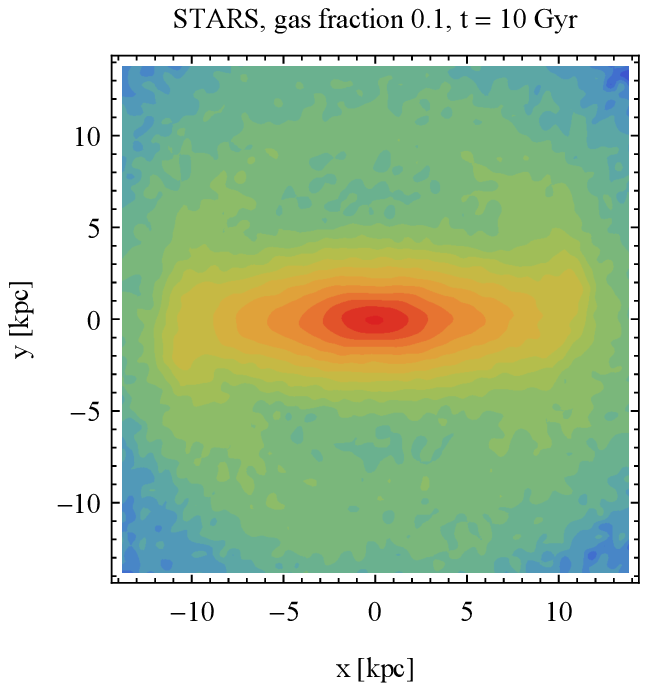}
\includegraphics[width=4.4cm]{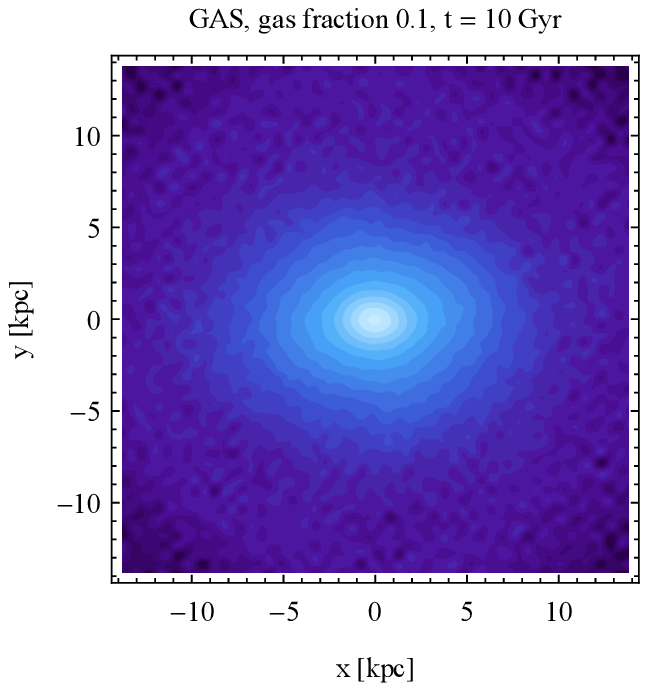} \\
\vspace{0.3cm}
\includegraphics[width=4.4cm]{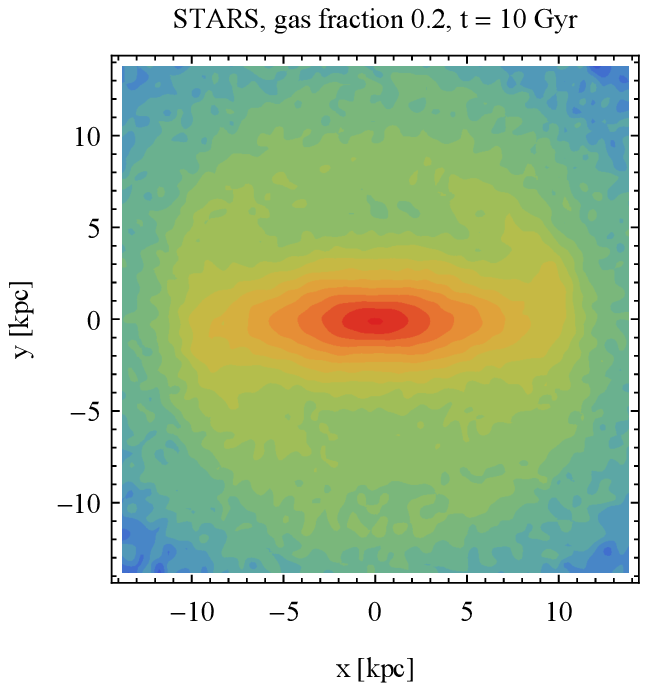}
\includegraphics[width=4.4cm]{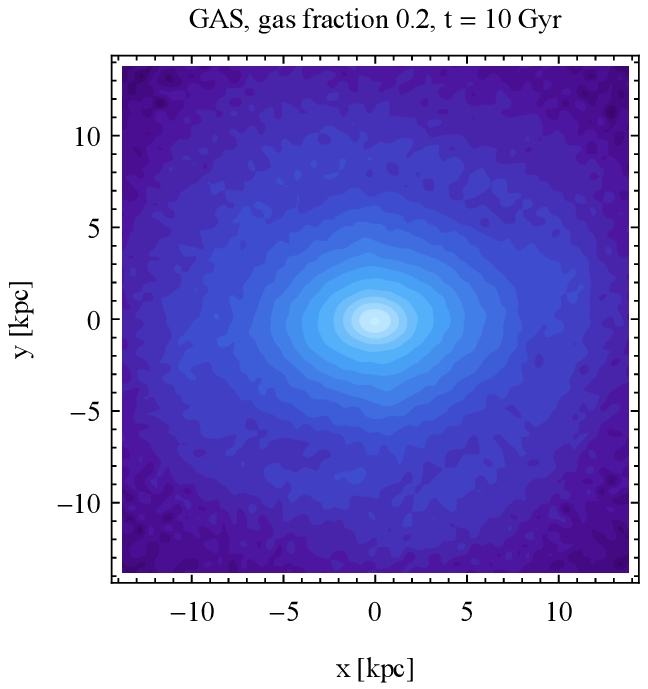} \\
\vspace{0.3cm}
\includegraphics[width=4.4cm]{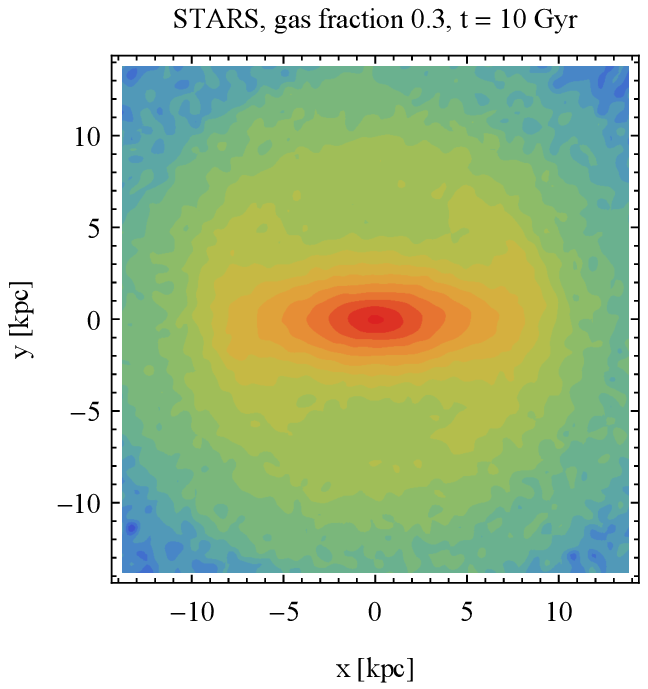}
\includegraphics[width=4.4cm]{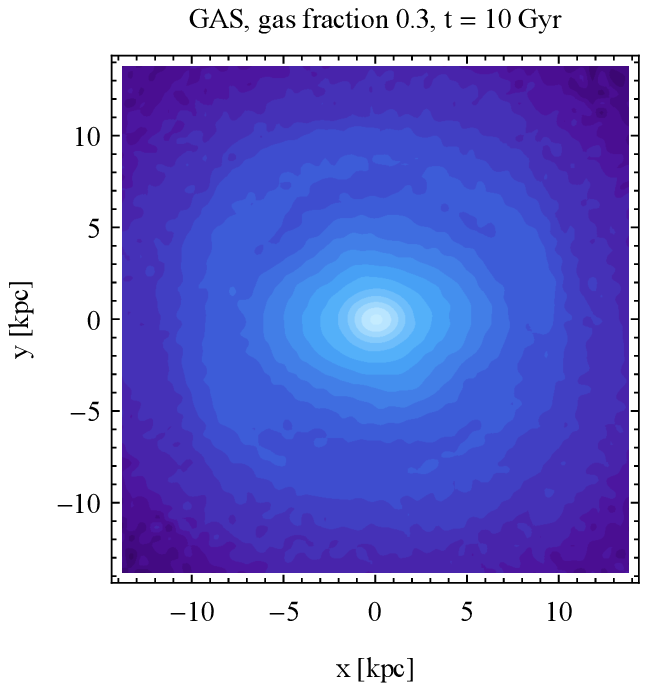}
\caption{Same as Fig.~\ref{surdenxy4gyr}, but at the end of the simulations after 10 Gyr.}
\label{surdenxy10gyr}
\end{figure}

The evolution of the galaxies was followed for 10 Gyr with the GIZMO code \citep{Hopkins2015}, an extension
of the widely used GADGET-2 code \citep{Springel2001, Springel2005}, saving outputs every 0.05 Gyr. The hydrodynamics
was solved using the Lagrangian Meshless Finite-Mass (MFM) method, assuming the polytropic equation of state
($\gamma=5/3$) for the gas. In order to keep the gas fraction constant in each simulation for the entire time, the gas
was not allowed to cool or form stars so no prescriptions for cooling, star formation, or feedback were used. The
softening scales were chosen following the recommendations in \citet{Hopkins2018} and they were fixed with
$\epsilon_{\rm D} = 12$ pc and $\epsilon_{\rm H} = 75$ pc for the stellar disk and halo of the galaxy, respectively.
For the gas, we used an adaptive gravitational softening with the minimum value of $\epsilon_{\rm G} = 1$ pc.

\begin{figure}
\centering
\includegraphics[width=9cm]{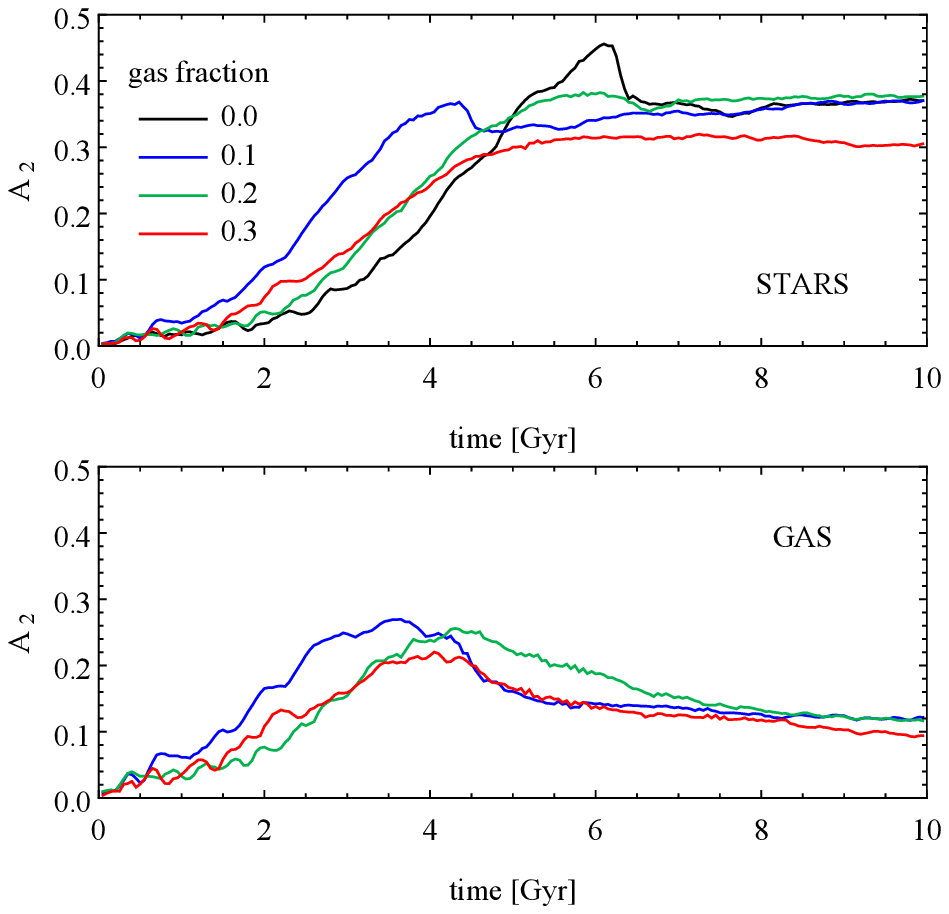}
\caption{Evolution of the bar mode $A_2$ for the stellar component (upper panel) and the gas (lower panel). The
measurements were performed within $2 R_{\rm D}$, that is, the bar mode here is a single value
$A_2 = A_2 (< 2 R_{\rm D})$. Lines of different colors correspond to different gas fractions.}
\label{a2}
\end{figure}

\section{Evolution of the bars}

The galaxies with a gas fraction of 0.4 and higher did not form a bar at all, so in the following analysis we restrict
the discussion to the four simulations with gas fractions of $f=0, 0.1, 0.2,$ and 0.3. The face-on surface density
distributions of the stars and gas in the disks at 4 and 10 Gyr are shown in Fig.~\ref{surdenxy4gyr} and
\ref{surdenxy10gyr}, respectively. In each of the figures, the plots in the left column are for the stars and those in
the right one are for the gas, while subsequent rows correspond to an increasing gas fraction. We see that after 4 Gyr
of evolution (Fig.~\ref{surdenxy4gyr}), the bars are already quite strong, both in the stellar and gaseous component.
On the other hand, at the end of the simulations, after 10 Gyr (Fig.~\ref{surdenxy10gyr}), while the stellar bars
become even stronger, the ones in the gas dissolve and only leave behind slightly oval shapes.

The simplest way to quantify the formation and evolution of the bar is to measure its strength as the $m=2$ mode of the
Fourier decomposition of the surface distribution of particles projected along the short axis: $A_m (R) = | \Sigma_j
\exp(i m \theta_j) |/N$. Here, $\theta_j$ is the azimuthal angle of the $j$th star and the sum is up to the total
number of $N$ particles. The radius $R$ is the standard radius in cylindrical coordinates in the plane of the disk, $R
= (x^2 + y^2)^{1/2}$. Figure~\ref{a2} shows this quantity measured for particles within twice the initial disk
scale-length, $2 R_{\rm D}$. The upper panel of the figure shows the strength of the bar in stars and the lower one for
the gas, with colors coding the gas fraction. For the gas fraction $f=0.4$ $A_2$ is always below 0.06, both for stars
and gas; we do not show this in the figure. We note that there is no specific trend in the stellar bar formation time,
that is, the galaxy with $f=0.1$ forms the bar first. This may be due to particular realizations of the initial
conditions, which would be an example of stochasticity that is associated with bar formation and evolution
\citep{Sellwood2009}, although the presence of the gas may also play a role, as discussed below.

Figure~\ref{a2} demonstrates that the stellar bar is the strongest in the purely stellar disk ($f=0$) and the weakest
in the high gas fraction case ($f=0.3$).\ However, for the intermediate gas fractions ($f=0.1$ and 0.2), the maximum
strength of the bar is similar so there is no obvious monotonic dependence of the bar strength on the gas fraction. In
addition, while the disks with gas fractions of $f=0, 0.2,$ and 0.3 reach the maximum bar strength after about 6 Gyr of
evolution, this maximum is reached much earlier (around 4.3 Gyr) for the low gas fraction case ($f=0.1$). In spite of
this, all stellar bars seem to end up with a similar strength at the end of the simulations, after 10 Gyr, except for
the $f=0.3$ case, which has a slightly weaker bar.

\begin{figure*}
\centering
\includegraphics[width=7cm]{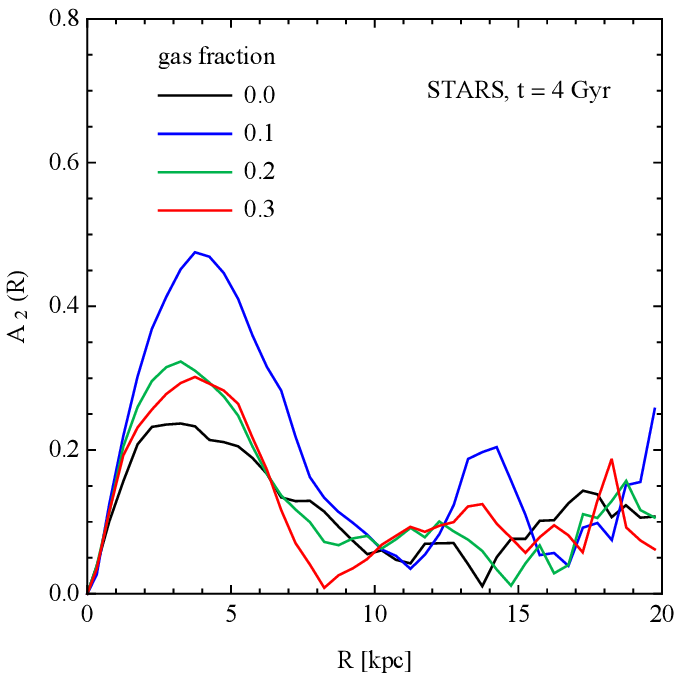}
\includegraphics[width=7cm]{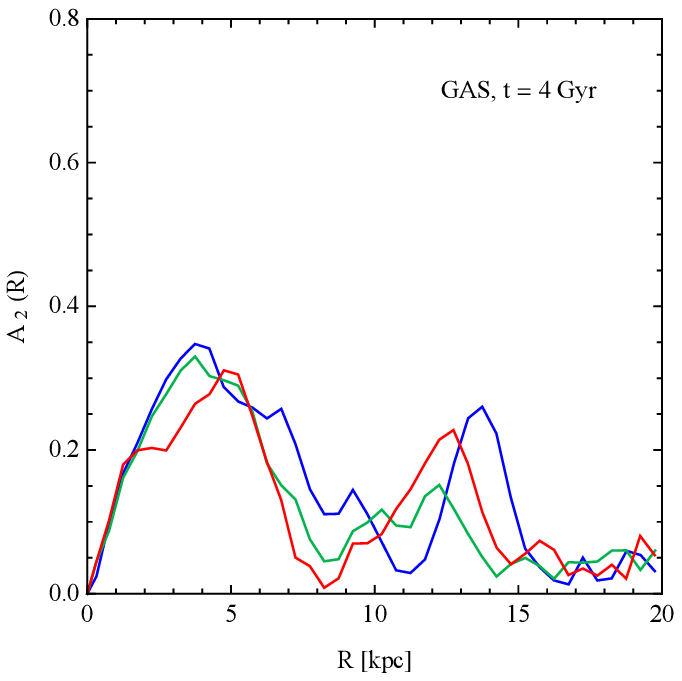} \\
\includegraphics[width=7cm]{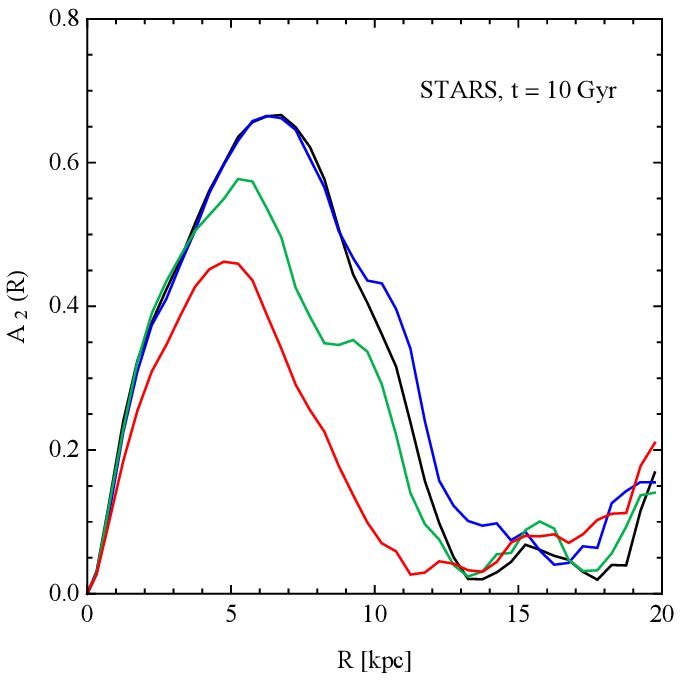}
\includegraphics[width=7cm]{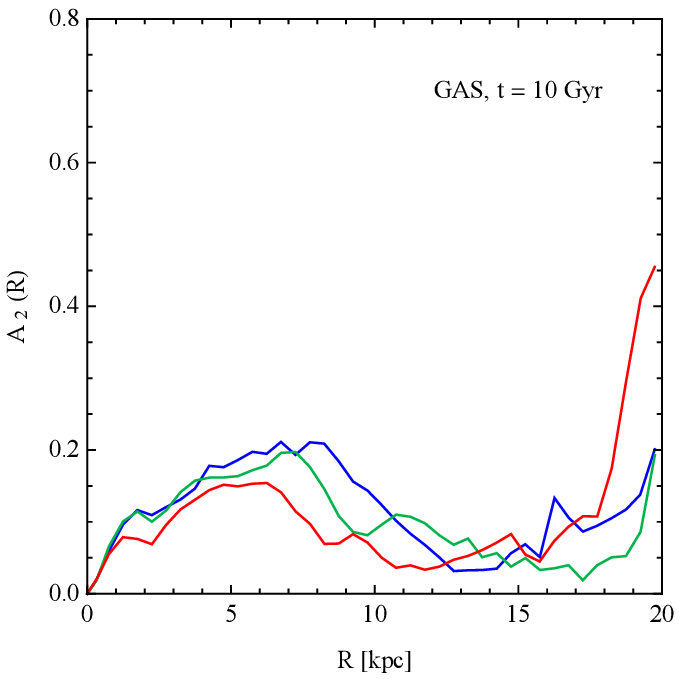}
\caption{Profiles of the bar mode $A_2 (R)$ for the stars (left column plots) and gas (right column). Rows
correspond to different stages of evolution: after 4 Gyr (upper row) and after 10 Gyr (lower row). Lines of different
colors show results for different gas fractions.}
\label{a2profiles}
\end{figure*}

\begin{figure*}
\includegraphics[width=8.9cm]{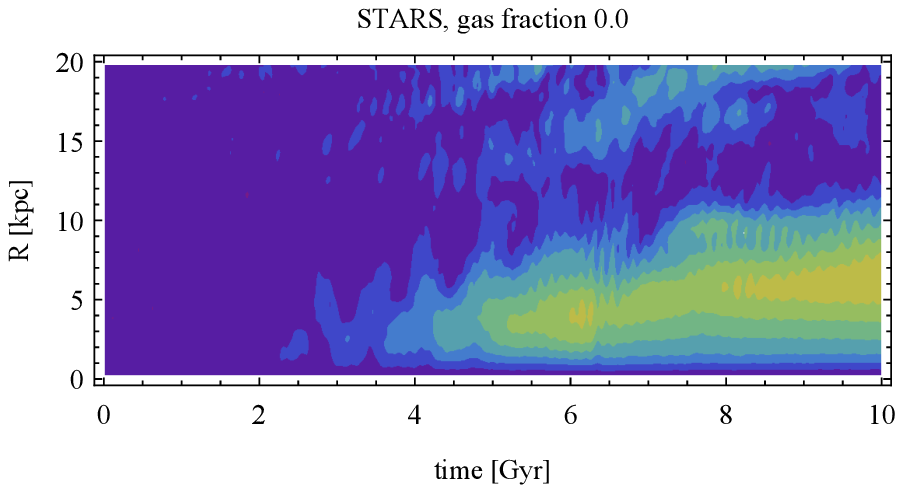}
\hspace{2.5cm} \includegraphics[width=4cm, bb= 3.0705cm -2.48cm 8.22cm -1.35cm]{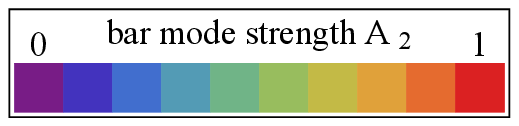} \hspace{6cm}  \\
\vspace{0.3cm}
\hspace{-0.18cm}
\includegraphics[width=8.9cm]{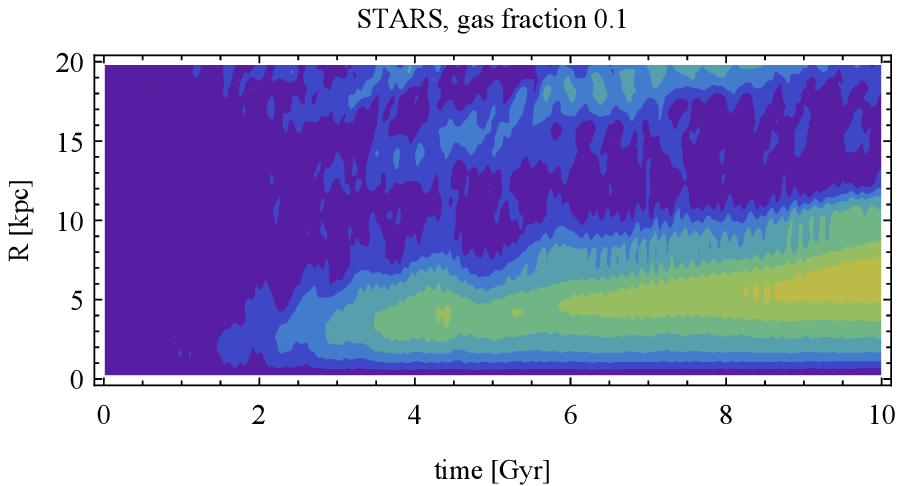}
\includegraphics[width=8.9cm]{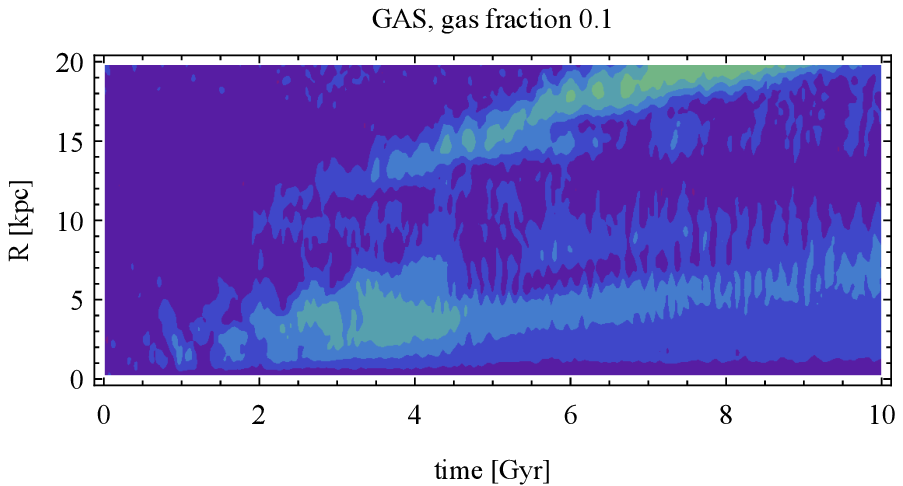} \\
\vspace{0.3cm}
\hspace{-0.18cm}
\includegraphics[width=8.9cm]{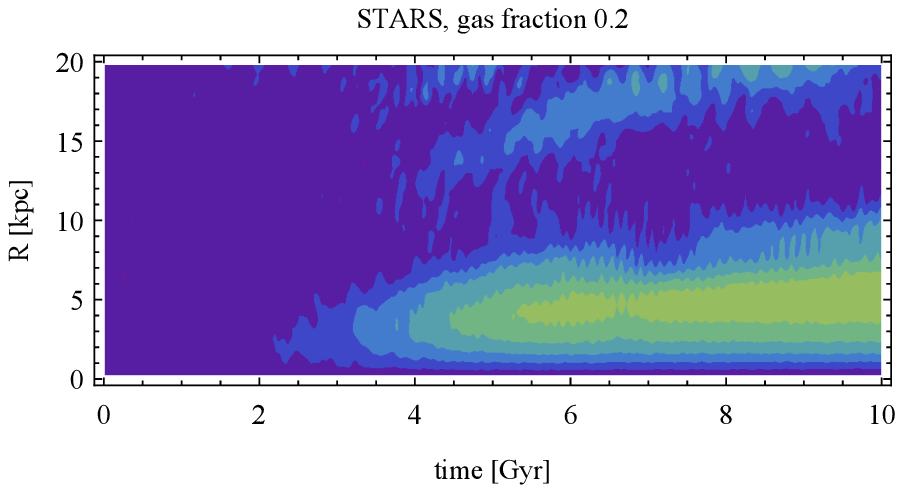}
\includegraphics[width=8.9cm]{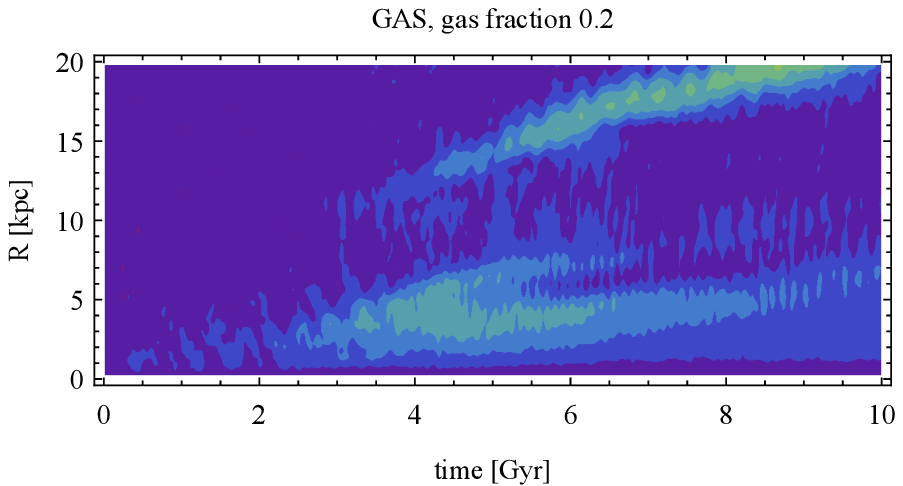} \\
\vspace{0.3cm}
\hspace{-0.18cm}
\includegraphics[width=8.9cm]{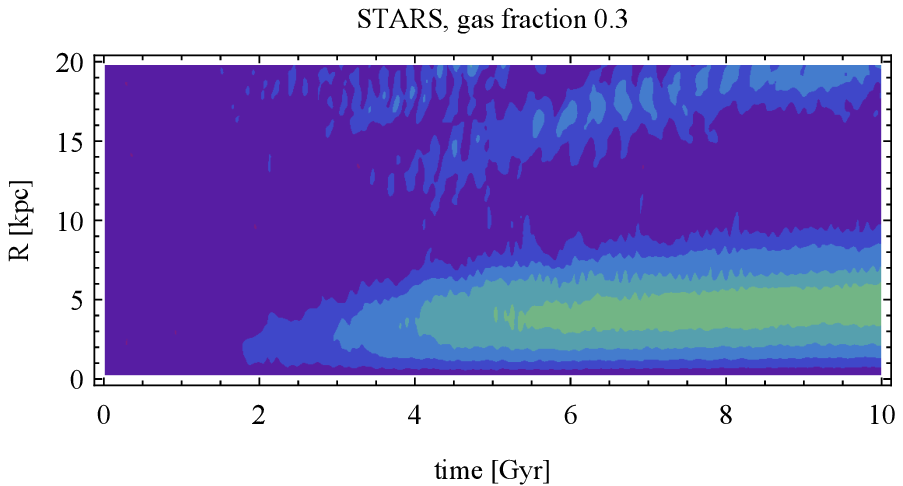}
\includegraphics[width=8.9cm]{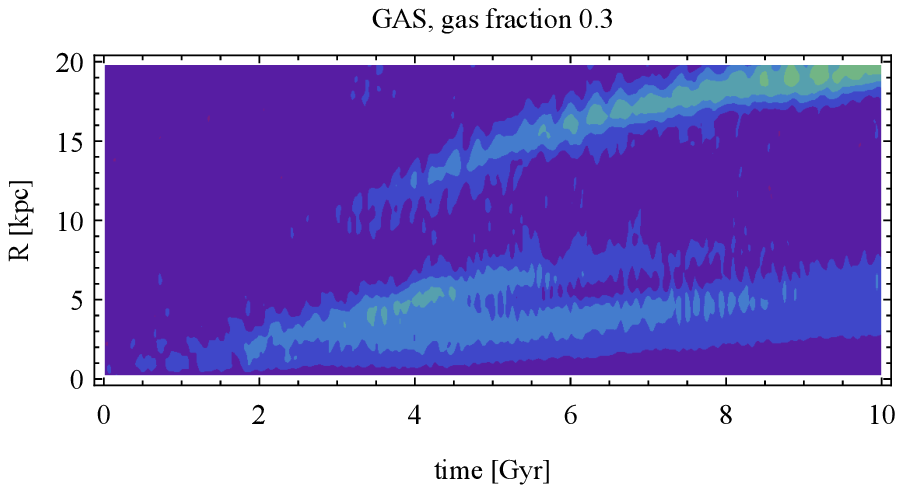}
\caption{Evolution of the bar mode profiles $A_2(R)$ in time. The left column plots show the results for stars
and the right ones for the gas. Rows from top to bottom correspond to increasing gas fraction in the galaxies.}
\label{a2modestime}
\end{figure*}

All the bars with a gas fraction below 0.3 show a monotonic growth until they reach a maximum, after which the bar mode
suddenly drops. This abrupt decrease of the bar strength is a clear signature of the buckling instability. We note that
this drop is more pronounced the lower the gas fraction and it does not occur at all for $f=0.3$. This suggests that
buckling does not happen in our highest gas fraction case.

The evolution of the bars in the gaseous component is significantly different, as demonstrated by the lower panel of
Fig.~\ref{a2}. Again, the bar for $f=0.1$ grows the fastest, but all the bars reach maximum strength around 4 Gyr after
which time they start to decline slowly, while the stellar bars still grow or at least regain their strength after
buckling. We also note that the time when the gaseous bars start to decline is not related to the time of
buckling of the stellar bar. While for $f=0.1$ these times coincide, for $f=0.2$ the gas bar starts to decline well
before the stellar bar buckles, and for $f=0.3$ the gas bar declines despite the fact that the stellar bar
does not seem to buckle at all.

A more detailed picture of the bar strength and length can be obtained from measurements of the profiles of the bar
mode, as a function of the cylindrical radius, $A_2 (R)$. A few examples of such profiles are shown in
Fig.~\ref{a2profiles}. In rows, we plotted the profiles at different times, $t=4$ and 10 Gyr, and in columns the
results for stars (left column) and gas (right column) are shown. These measurements confirm that at 4 Gyr,
the stellar bar is indeed the strongest for the galaxy with a gas fraction of $f=0.1$ and the gas bars are similar in
all simulations. Instead, at the end of the evolution,  the stellar bar is almost identical for $f=0$ and $f=0.1$, a
little weaker for $f=0.2,$ and even more so for $f=0.3$. The gas bars are almost gone by then since $A_2 (R)$ do not
exceed 0.2 at any radius, but they are still a bit stronger and longer for lower gas fractions.

The profiles shown in Fig.~\ref{a2profiles} can also be used to estimate the length of the bars. A common way to do
this is to estimate the radius where the $A_2 (R)$ value drops to half the maximum of the profile. We thus note that at
the end of the evolution, the bars in both components are significantly shorter for higher gas fractions.

A full picture of the bar evolution is presented in Fig.~\ref{a2modestime}, which combines the dependence of $A_2$ on
both the radius and time. All stellar bars start to grow around 2 Gyr, in the sense that around that time their $A_2$
profiles cross the threshold of 0.1 for the first time, although this happens a little earlier for the galaxy
with a gas fraction of $f=0.1$. Interestingly, small perturbations in the gas component tend to appear earlier, already
around $t=0.5$ Gyr in all cases. They increase particularly fast in the $f=0.1$ case and may provide the seed for the
stellar bar, thus explaining its faster growth in this simulation. The effect seems to also be present for the higher
gas fractions $f=0.2$ and 0.3 since these bars also start to form earlier than the one with $f=0$, although the
fluctuations in the gas distribution are weaker than for $f=0.1$.

Signatures of buckling are also clearly present in these images. The stellar bars grow in strength and length until
they suddenly weaken and shorten due to buckling. This happens around 6.3, 4.5, and 6.7 Gyr for $f=0$, 0.1, and 0.2
respectively, so this is significantly earlier for $f=0.1$ with respect to the other two cases as a result of faster
bar growth in this case. However, no such event is visible for the largest gas fraction $f=0.3$. This bar grows
steadily until about 6 Gyr and then remains approximately the same until the end of evolution. Instead, the stellar
bars in the lower gas fraction cases all start to grow again in length and strength after the buckling event.

One more difference between the stellar and gaseous bars, in addition to their different lifetimes, is their shape. As
can be seen in Figs.~\ref{a2profiles} and \ref{a2modestime} for the gas component, the profiles $A_2(R)$ grow much more
slowly in radius compared to the stars. This is also seen in the images of the bars shown in Fig.~\ref{surdenxy4gyr}:
the contours of the gas bars, even at their strongest, are much more circular in the center than those in the stellar
bars, which are elongated. They become more and more circular in time at larger radii, leading to the transformation
into oval shapes rather than bars at the end of evolution.

As far as the pattern speeds of the bars are concerned, they are similar in all cases: for stellar and gas bars and in
simulations with different gas fractions. Their values are always around $\Omega_{\rm p} = 20$ km s$^{-1}$ kpc$^{-1}$
at the time of bar formation (defined as the moment of crossing the threshold of $A_2(R)=0.2$ for the first time) and
they decrease down to about 10 km s$^{-1}$ kpc$^{-1}$ at 10 Gyr.

Another pronounced morphological feature present in the images of Fig.~\ref{a2modestime} beside the bar
is the branch of increased $A_2(t, R)$ values extending from the time-radius position $(t, R) =$ (4 Gyr, 12 kpc) to (10
Gyr, 20 kpc), which is present in all panels at approximately the same location and is even stronger in gas than in
stars. This feature reflects the presence of an almost ring-like, two-armed spiral structure with a very small pitch
angle, forming at the corotation radius. Its increasing radius is due to the fact that while the mass distribution in
the galaxies does not strongly evolve with time, the pattern speed of the bar decreases. The corotation radius, where
the circular frequency equals the pattern speed, $\Omega = \Omega_{\rm p}$, thus increases with time from around 10 kpc
at the time of the formation of the bar up to 20 kpc at 10 Gyr.

\begin{figure}
\centering
\includegraphics[width=9cm]{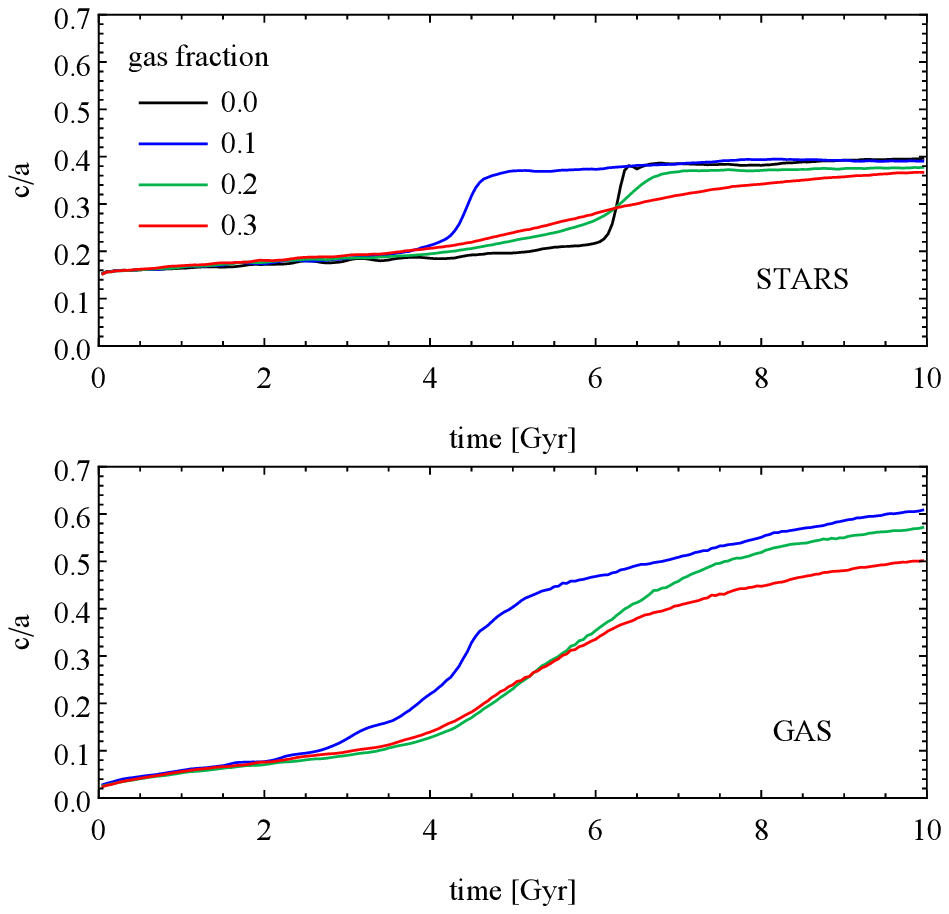}
\caption{Evolution of the ratio of the shortest to longest axis (the vertical thickness to the
radial size of the bar) for the stars (upper panel) and gas (lower panel).
Measurements were performed for particles inside $2 R_{\rm D}$. Lines of different
colors correspond to different gas fractions.}
\label{ca}
\end{figure}

\begin{figure}
\centering
\includegraphics[width=9cm]{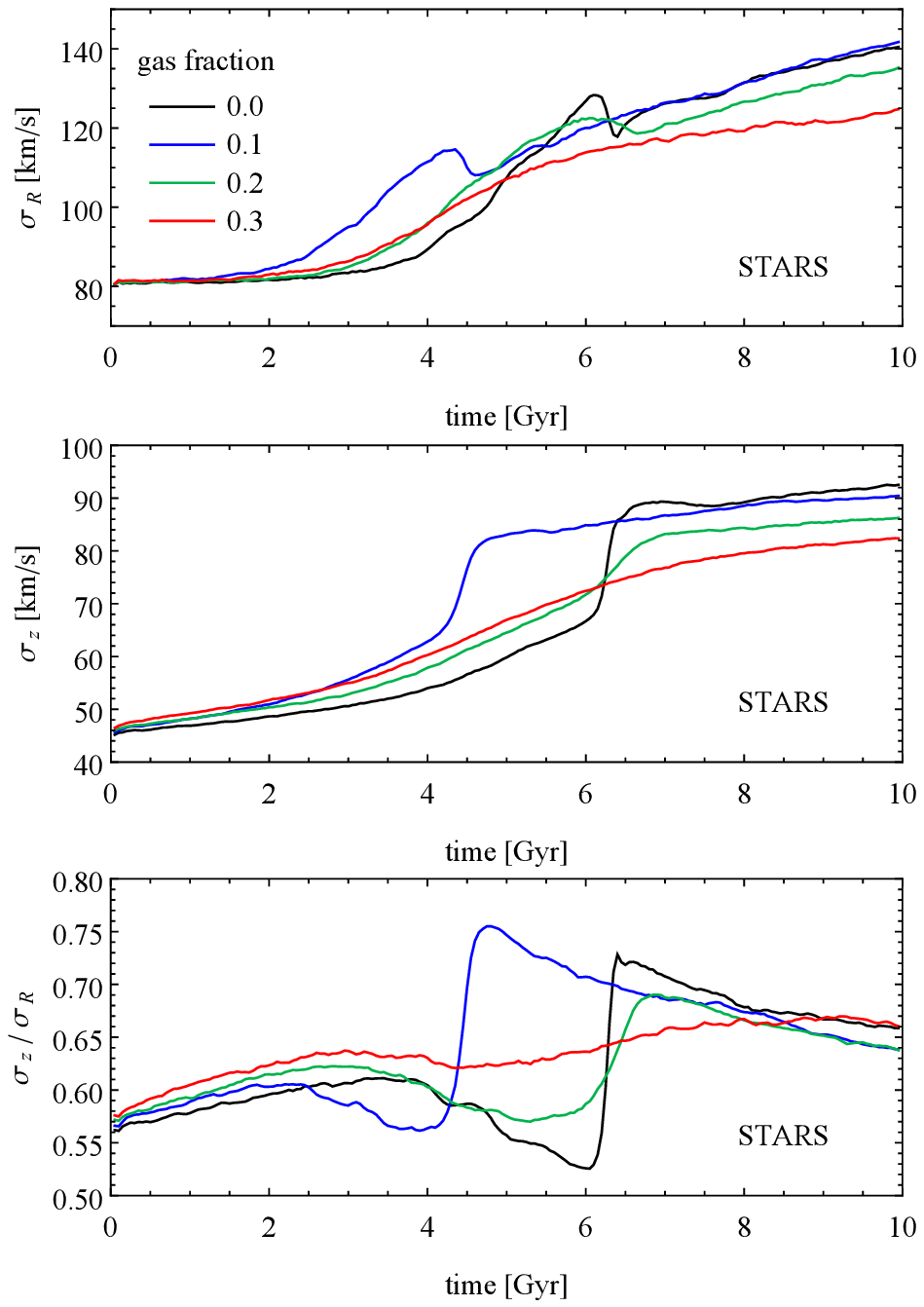}
\caption{Evolution of the kinematics of the stars in the form of the radial velocity dispersion (upper panel), vertical
velocity dispersion (middle panel), and the ratio of the two quantities (lower panel). Measurements were performed for
stars inside $2 R_{\rm D}$. Lines of different colors correspond to simulations with different gas fractions.}
\label{kinematics}
\end{figure}

\section{Buckling instability}

In addition to the telltale signatures in the form of the weakening of the bar, the presence and effects of buckling
can also be detected by measurements of the evolution of the shape of the stellar or gaseous component and the
kinematics of the stars. Buckling is expected to thicken the bar significantly and this is indeed what we see in
Fig.~\ref{ca}, showing the evolution of the ratio of the shortest to longest axis $c/a$ of the distribution of stars
and gas, which is a measure of the ratio of the vertical thickness to the radial size of the bar. For gas fractions
$f=0$ and 0.1, the values of $c/a$ for the stars (upper panel) show a sharp increase at the same times as when the bars
were weakened, which we identified as the times of buckling events. The increase is also present for $f=0.2,$ but it is
much milder. For $f=0.3,$ the increase is very slow and no specific upturn is seen. The gas thickens
even more, and this occurs at around the same time as for the stars. However, the thickening in the
gas happens more slowly in time, so it seems as though the gas is just responding to what is happening to the
stars, rather than buckling itself.

\begin{figure}[ht!]
\centering
\hspace{0.7cm} \includegraphics[width=6cm]{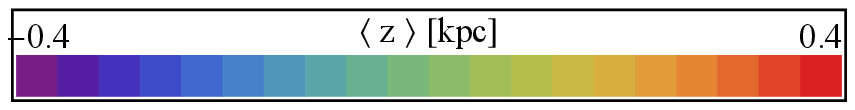} \\
\vspace{0.3cm}
\includegraphics[width=8.9cm]{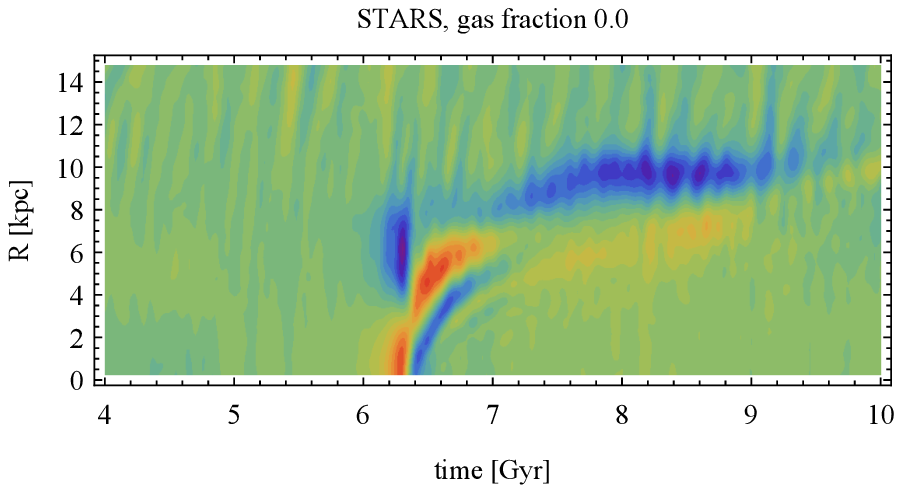} \\
\vspace{0.3cm}
\includegraphics[width=8.9cm]{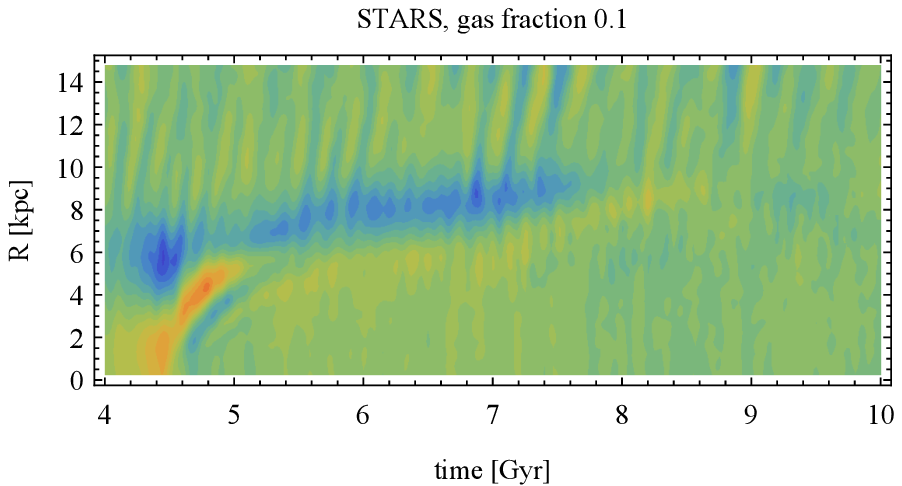} \\
\vspace{0.3cm}
\includegraphics[width=8.9cm]{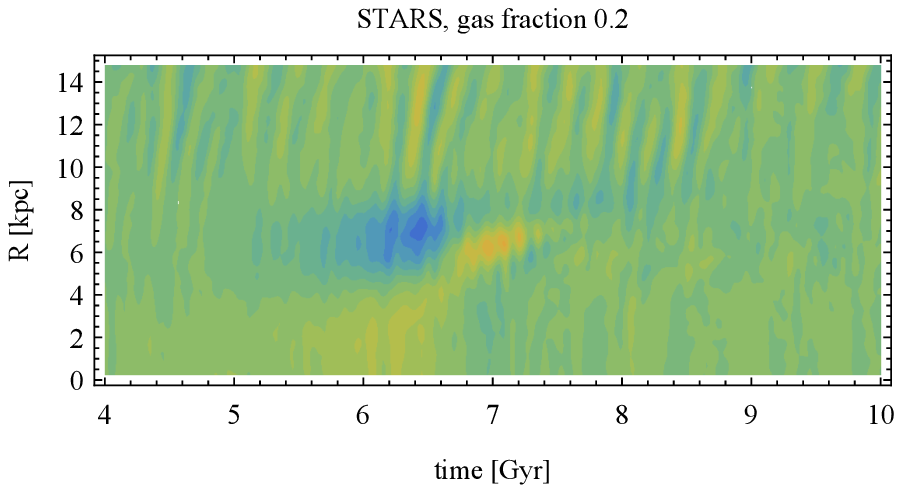} \\
\vspace{0.3cm}
\includegraphics[width=8.9cm]{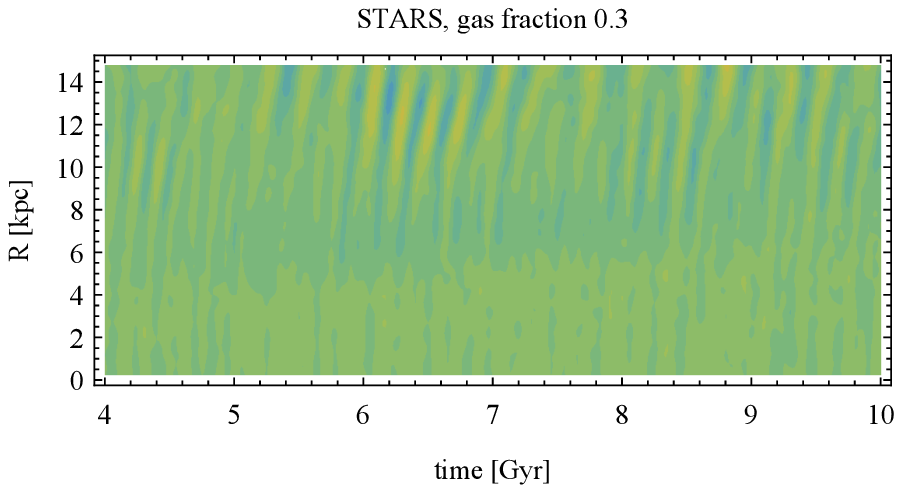}
\caption{Evolution of the profiles of the mean distortion of
the positions of the stars along the vertical axis $z$ in simulations with different gas fractions (from top to
bottom). Positive quantities point along the angular momentum vector of the disk.}
\label{meanzprofiles}
\end{figure}

\begin{figure}[h!]
\centering
\includegraphics[width=4.4cm]{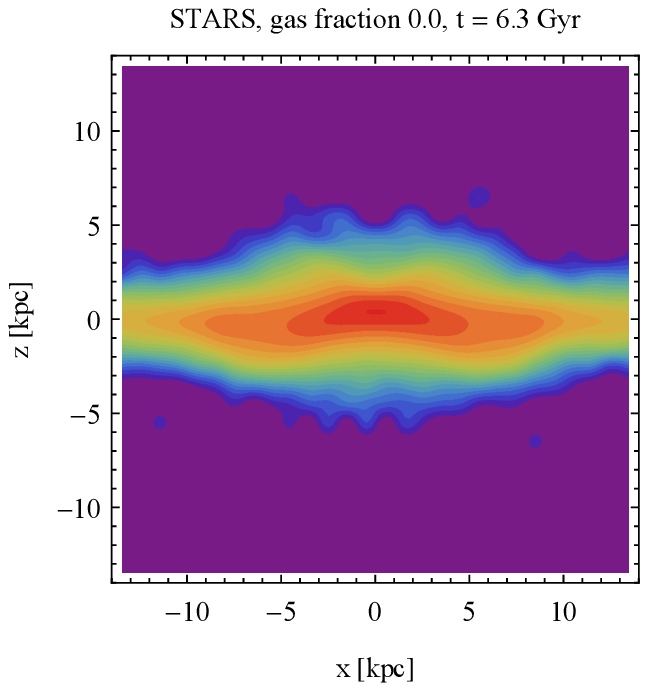}
\includegraphics[width=4.4cm]{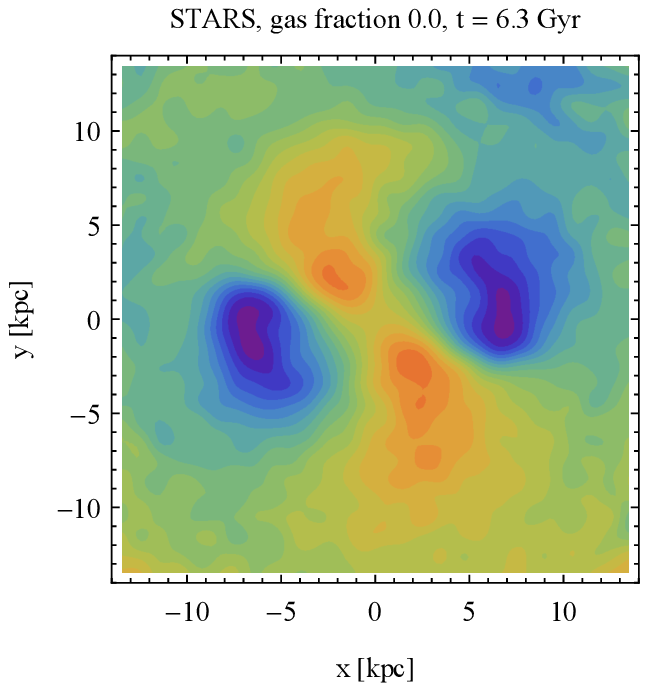}\\
\vspace{0.3cm}
\includegraphics[width=4.4cm]{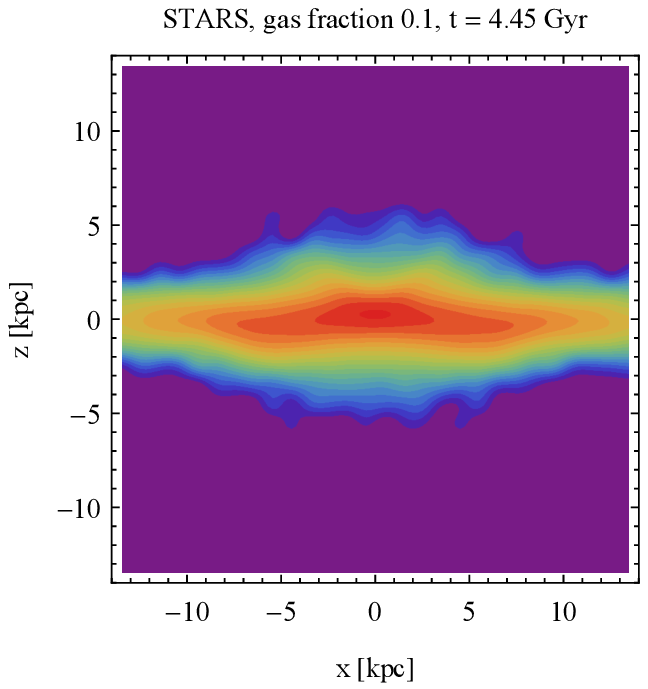}
\includegraphics[width=4.4cm]{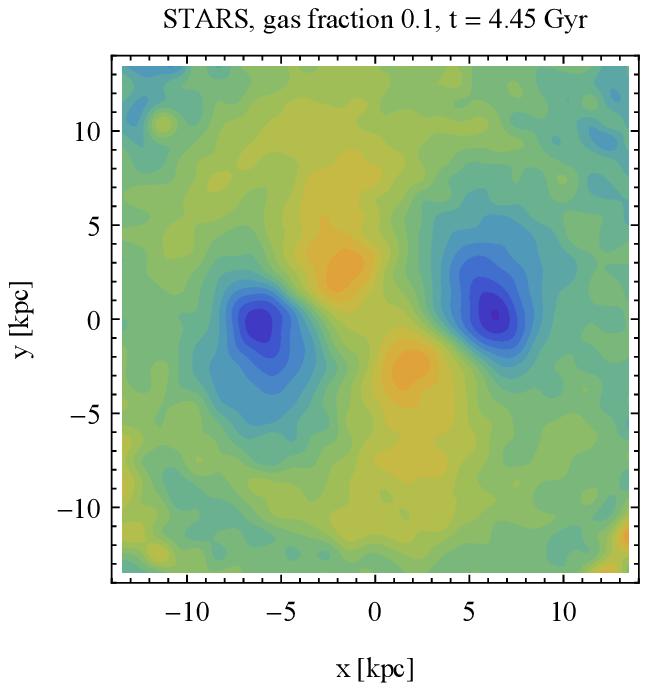}\\
\vspace{0.3cm}
\includegraphics[width=4.4cm]{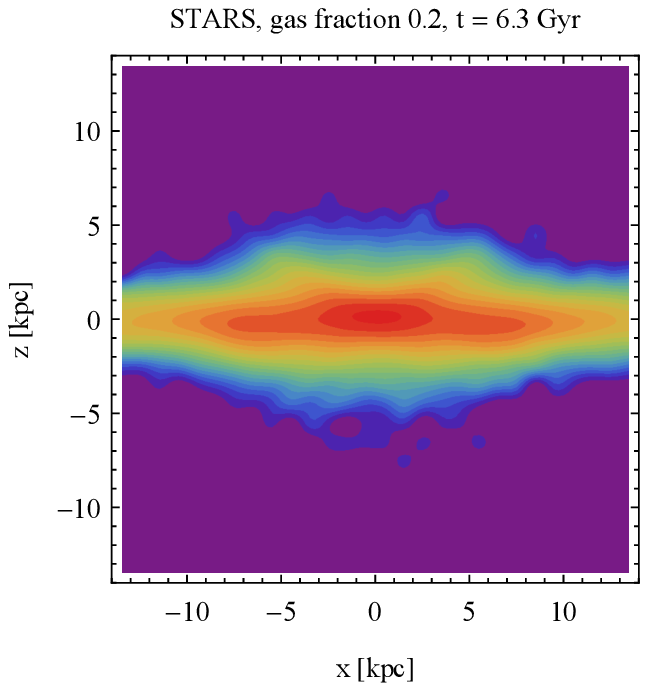}
\includegraphics[width=4.4cm]{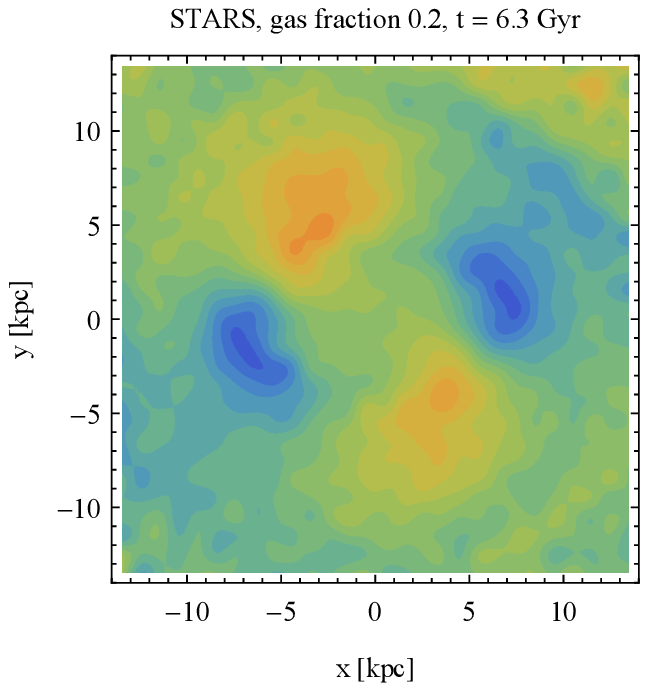}\\
\vspace{0.3cm}
\includegraphics[width=4.4cm]{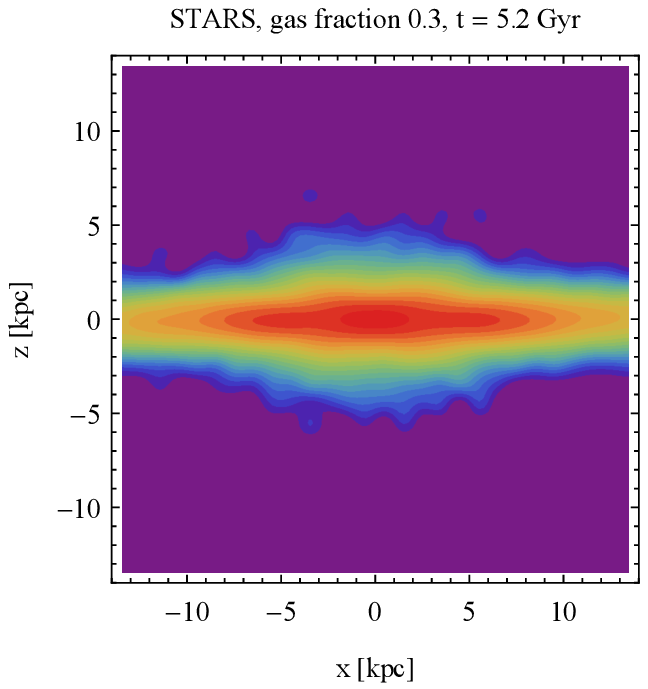}
\includegraphics[width=4.4cm]{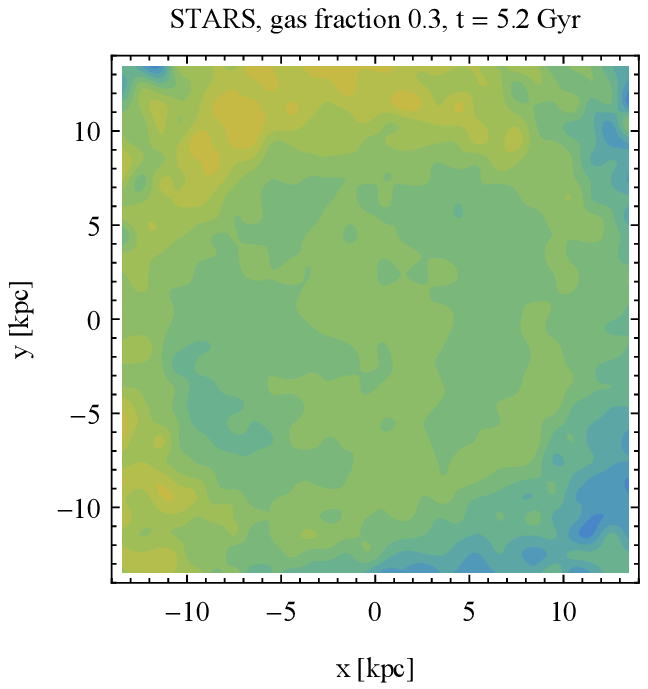}\\
\vspace{0.2cm}
\hspace{0.61cm}
\includegraphics[width=3.69cm]{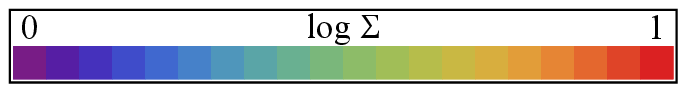}
\hspace{0.61cm}
\includegraphics[width=3.69cm]{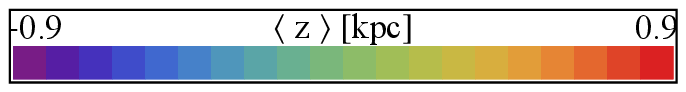}\\
\vspace{0.3cm}
\caption{Distortion of the stellar component during the first phase of buckling. Left column: surface density
distributions of the stellar component viewed edge-on. The surface density was normalized to
the central maximum value in each case. Right column: the face-on
maps of the mean distortion out of the disk plane along the vertical direction. Positive quantities
point along the angular momentum vector of the disk. Rows show the results for simulations
with different gas fractions.}
\label{surdenzmap}
\end{figure}

\begin{figure}
\centering
\includegraphics[width=7cm]{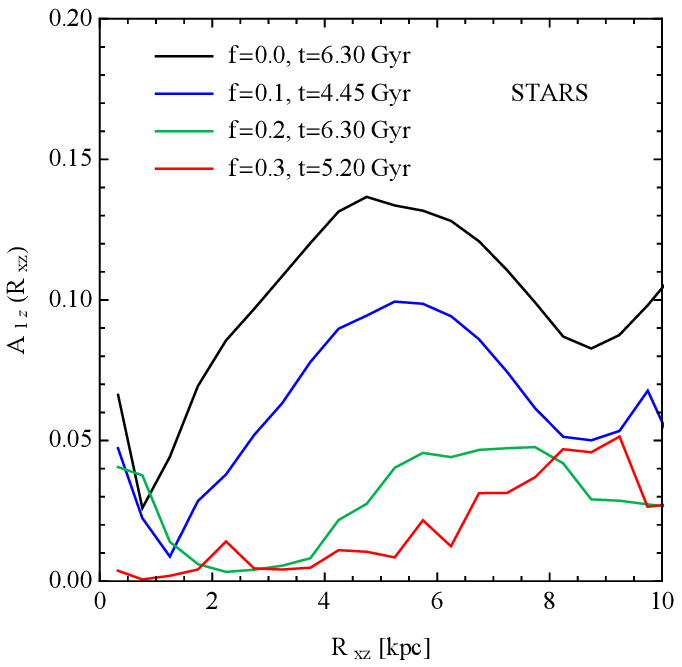}
\caption{Profiles of the asymmetry measure $A_{1z}$ of the stellar component
in the edge-on view during buckling.}
\label{a1profiles}
\end{figure}

The evolution of the kinematics of the stars confirms this picture. Figure~\ref{kinematics} shows the velocity
dispersion of the stars along the cylindrical radius (upper panel), along the vertical direction (middle panel), and the
ratio of the two (lower panel). While the evolution of the radial velocity dispersion reflects the evolution of the bar
strength (the stronger the bar, the more radial the stellar orbits or the more stars on radial orbits), the
vertical dispersion detects buckling by measuring the increase of stellar velocities in the direction perpendicular to
the disk. As in the case of thickness, the simulations with low gas fractions show a sharp increase in this dispersion
in contrast to the cases with higher gas fractions. The effect is even more pronounced in the evolution of the ratio of
the two dispersions. We note that the minimum value of the ratio just before buckling is different in different
simulations, so it is unlikely that the instability is triggered by the ratio decreasing below some
characteristic threshold as required by the interpretation of buckling based on the fire-hose instability.

So far we have only demonstrated that, except for the case with $f=0.3$, all our bars are weakened and thickened and
their stars acquire vertical motions. However, the buckling instability is also supposed to induce temporary
distortions of the bar out of the disk plane. In order to detect such a distortion, we measured the mean position of
the stars along the vertical direction as a function of the cylindrical radius. Figure~\ref{meanzprofiles} shows these
measurements for the stellar components of the galaxies for the periods between 4 and 10 Gyr, that is, when the bar is
already formed in all cases. We see that the strength of the buckling depends very much on the gas fraction in the
disk. The times when the distortions occur coincide with the times of sudden weakening and thickening of the bars
discussed above, thus confirming that buckling was responsible for these events.

The distortion is clearly the strongest for the purely collisionless case with no gas (the upper panel of
Fig.~\ref{meanzprofiles}). Its shape changes from frown- to smile-like in the first phase of buckling, around 6.3 Gyr.
After that, the distortion propagates outward and disappears around 7 Gyr, to reappear again a bit later in
the form of secondary buckling, which lasts much longer (between 7.2 and 9 Gyr) and takes place in the outer part of
the bar ($R > 5$ kpc).

For the simulation with a gas fraction of $f=0.1$ (the second panel of Fig.~\ref{meanzprofiles}), the pattern is very
similar, including the presence of the secondary, longer buckling phase, although the distortion is
significantly weaker. It happens much earlier because the bar forms and grows earlier in this simulation. For the gas
fraction 0.2 (the third panel of the figure) the signal is even weaker and more extended in time, without any clear
evidence for the secondary buckling. For $f=0.3,$ a very weak distortion is present within $R < 8$ kpc, which remains
at the same level between 5 and 8 Gyr and does not increase, so there is no clear signature of buckling in this case.
For the gas component, the distortions, which are not shown here, are very weak and noisy even in the $f=0.1$ case and
they are not visible at all for higher gas fractions.

The character of the distortions is illustrated further in Fig.~\ref{surdenzmap} where we show in the left column the
edge-on surface density distributions of the stars and in the right one the face-on maps of the distortions out of the
disk plane. The rows correspond to the increasing gas fraction in the galaxies. The images are shown for the first
phase of buckling, when the inner part of the bar is the most strongly distorted upward and the outer part downward.
For each simulation, this occurs at a different time, so the outputs at $t=6.3$, 4.45, 6.3, and 5.2 were selected for
simulations with a gas fraction increasing from 0 to 0.3, respectively. We see again that the distortion weakens
considerably with the increasing gas content. Although the pattern has a similar shape in each case, even for $f=0.3$,
the distortion remains very weak in this case during the whole evolution. Similar patterns are also seen in the
velocity distribution, which is not shown here. The subsequent evolution of the patterns is similar to the one
discussed in detail in \citet{Lokas2019b}: they wind up and dissolve leaving behind bars with a distinct boxy/peanut
shape.

To further explore the nature of the distortions, especially in the doubtful case of $f=0.3,$ we calculated another
commonly used measure of the asymmetry in the form of the $m=1$ mode of the Fourier decomposition of the surface
distribution of stars in the edge-on view (projected along the intermediate $y$ axis): $A_{mz} (R_{xz}) = | \Sigma_j
\exp(i m \theta_j) |/N$. Here, $\theta_j$ is the azimuthal angle of the $j$th star and the sum is up to the total
number of $N$ stars, as for the $A_2$ calculation, but now the radius $R_{xz}$ is measured in the $xz$ plane, $R_{xz} =
(x^2 + z^2)^{1/2}$. The $A_{1z} (R_{xz})$ profiles for the times where the maximum distortions occur for different
simulations are shown in Fig.~\ref{a1profiles}. The profiles have a similar shape for $f=0$-0.2, with a maximum within
$2 R_{\rm D}$.\ While the shape for $f=0.3$ is different, it remains flat in the inner part of the galaxy and only
increases at larger radii. This behavior confirms that in this case the buckling event did not occur.

\begin{figure}
\centering
\includegraphics[width=4.4cm]{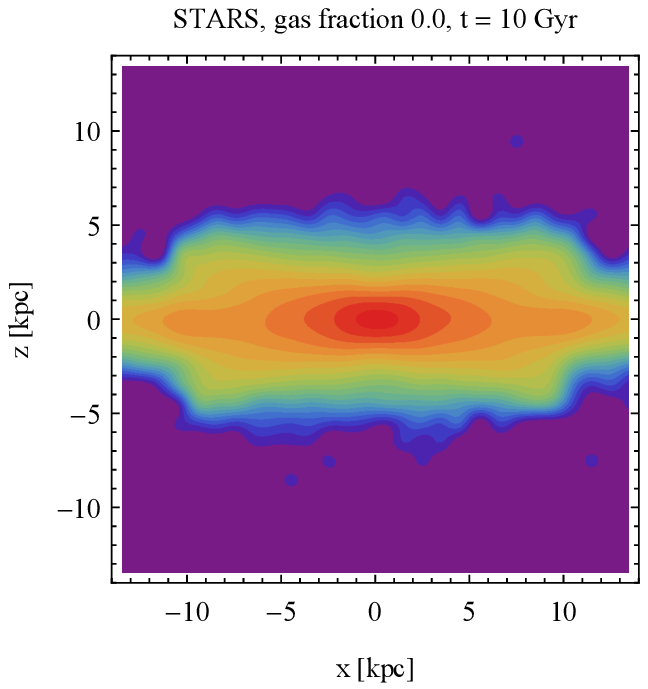}
\includegraphics[width=4.4cm]{legend19a.eps} \\
\vspace{0.3cm}
\includegraphics[width=4.4cm]{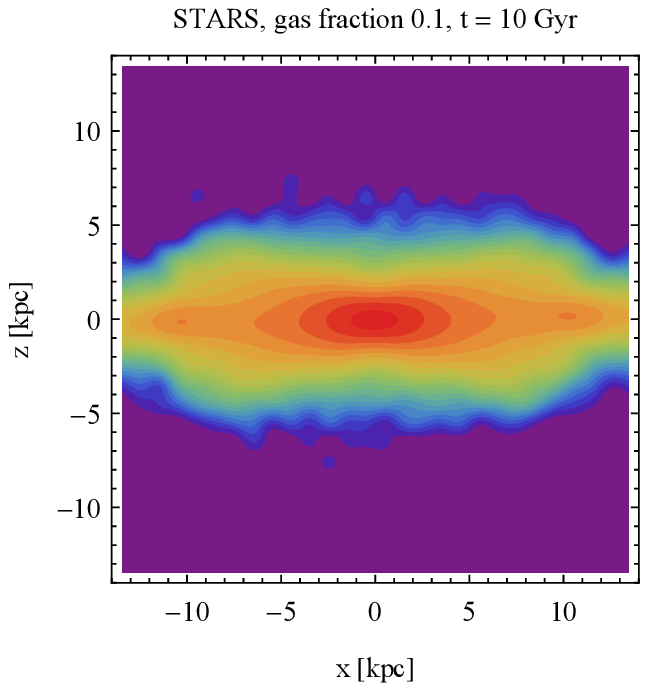}
\includegraphics[width=4.4cm]{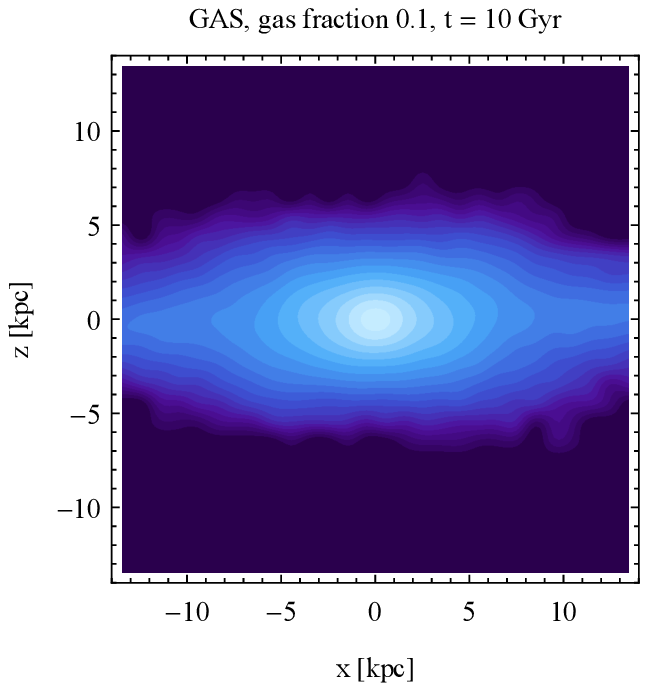} \\
\vspace{0.3cm}
\includegraphics[width=4.4cm]{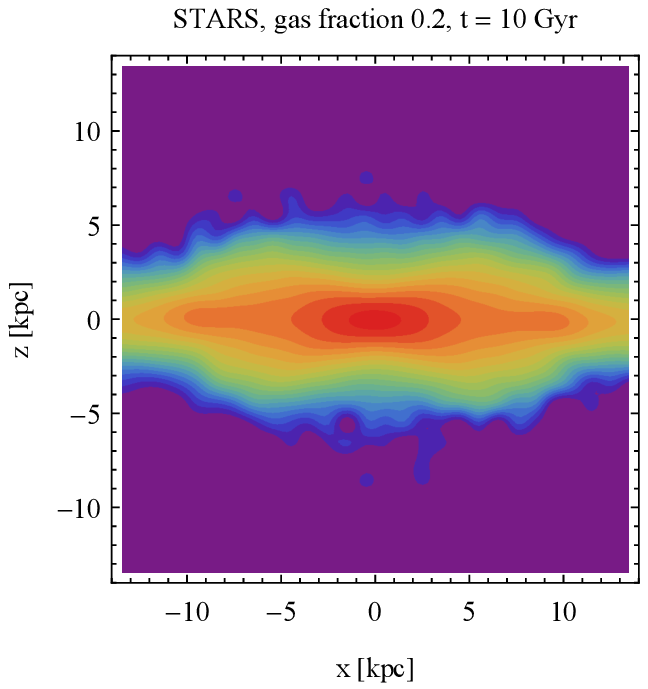}
\includegraphics[width=4.4cm]{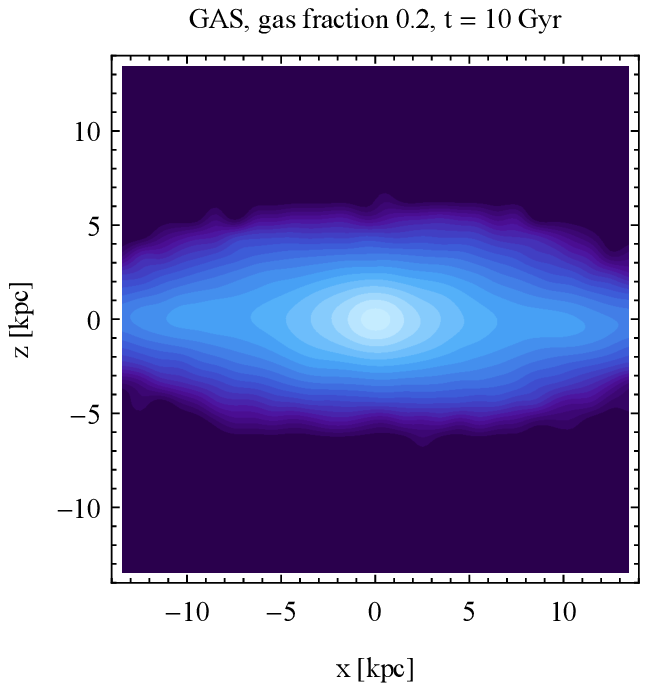} \\
\vspace{0.3cm}
\includegraphics[width=4.4cm]{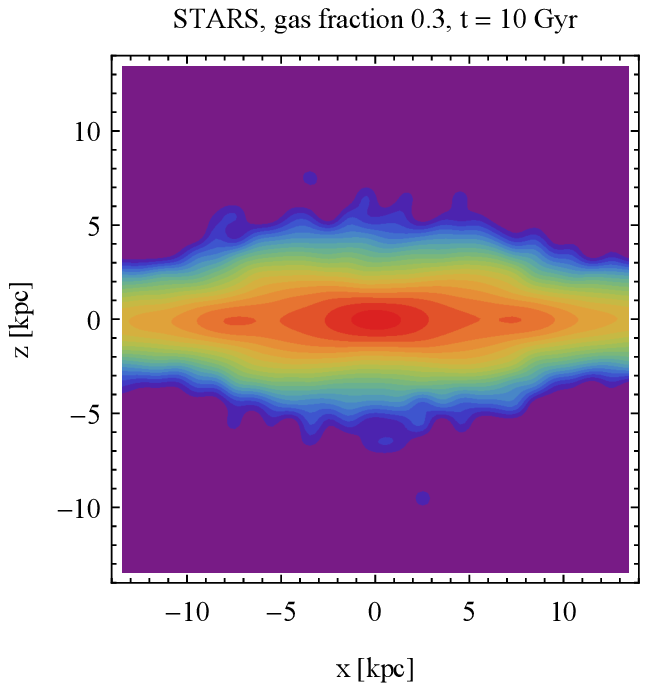}
\includegraphics[width=4.4cm]{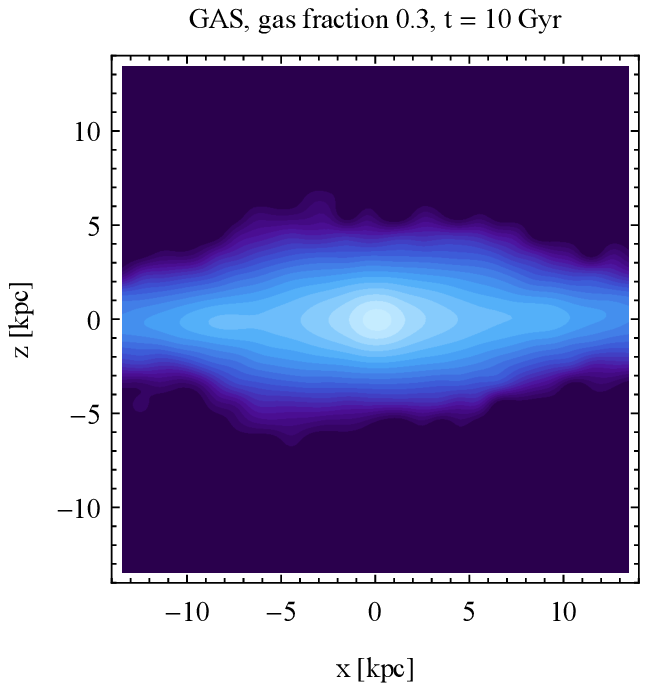}
\caption{Surface density distributions of stars and gas in the edge-on view at the end of the simulations, after 10
Gyr. Rows show the images from simulations with an increasing gas fraction. The left column is for stars and the right
one is for the gas. The surface density was normalized to the central value in each case.}
\label{surdenxz10gyr}
\end{figure}

In Fig.~\ref{surdenxz10gyr} we show the edge-on views of the galaxies at the end of evolution. The left-column plots
correspond to the stellar component and the right-column ones to the gas. Clear boxy/peanut density distributions are
present in stars as well as the gas, although their thickness decreases with an increasing gas fraction. This
is even the case for the galaxy with the highest gas fraction $f=0.3,$ which did not undergo buckling. Instead, the
stars and gas seem to have heated vertically with an outcome that is not qualitatively different from the bars that
underwent abrupt buckling events.

\section{Discussion}

We studied the formation and evolution of a bar in a galaxy with parameters similar to the Milky Way evolving in
isolation. The simulations differed only by the initial gas fraction, which was varied from $f=0$ to 0.3 with an
increment of 0.1. In order to isolate the effect of the warm gas on bar evolution, the gas fraction was kept constant
during the simulations, that is, the gas was not allowed to cool and form stars. Such a configuration may reflect, to
some extent, the real conditions where the gas that was already present in the disk is converted into stars,
but additional gas is constantly accreted onto the disk from the hot halo of the galaxy, thus keeping the gas
budget on the same level. Our main focus was to study, in detail, the effect of gas on the buckling instability in
galactic bars, which has so far been mainly discussed in the context of its effect on the longevity and survival of
bars.

In general, we find that bars forming in galaxies with gas are weaker than in galaxies without gas. For our particular
choice of initial galaxy parameters, for gas fractions higher than 0.3 the bar did not form at all. The exact value of
this threshold obviously depends on the disk susceptibility to bar formation, which is controlled by a number of
parameters including the disk mass and velocity dispersion. However, the trend of weaker bars for higher gas fractions
seems general and our results here agree with those of \citet{Athanassoula2013}. They find that for higher gas
fractions, the bars form later and grow more slowly. However, a direct comparison with our results is not possible
since in their simulations, the gas fractions strongly decrease with time due to star formation and the absence of any
external new gas supply to the disk. In our simulations, the bar formation is not slowed down by the presence of the
gas. An interesting case is that of the gas fraction $f=0.1$ where the bar formation is even speeded up, probably by
small fluctuations in the gas density which seed the stellar bar. But even though the stellar bar in this case has more
time to grow, it ends up with almost exactly the same strength and length as our bar in the purely collisionless
simulation (the lower left panel of Fig.~\ref{a2profiles}).

Our results also agree with recent studies of barred galaxies in cosmological simulations. In particular,
analyses of the Illustris and IllustrisTNG galaxy samples by \citet{Peschken2019} and \citet{Rosas2019}, respectively,
reveal that the fraction of barred galaxies is larger among gas-poor galaxies and that galaxies with stronger bars
tend to have less gas than their unbarred counterparts. We note, however, that \citet{Rosas2019} find strong bars even
in galaxies dominated by the gas component in the disk.

Although  the bars in the gas component start to grow earlier in general, we find that they are much weaker than the
stellar bars and reach their maximum strength around 4 Gyr. After that, they gradually weaken to acquire a spheroidal,
rather than a bar-like shape toward the end of the simulations. Thus we confirm that, in the words of
\citet{Binney2008} ``just as with humans, the dissipative lifestyle of a bar can lead to its early demise''. We
emphasize that this only applies to bars in the gas component. Once the stellar bars form, they survive until
the end of evolution in good shape; they are only weaker and shorter for higher gas fractions.

Weaker stellar bars that formed with a higher gas fraction experience weaker buckling instabilities. Their distortion
out of the disk plane is smaller and more extended in time. For the gas fraction $f=0.2,$ the secondary phase of
buckling is absent and for $f=0.3,$ the buckling does not occur. Still, the process of buckling happens in a similar
way in all cases: it begins with a quadrupole distortion of the bar when seen face-on, which is probably due to
banana-like orbits. This pattern later winds up in the form of kinematic bending waves, as described in
\citet{Lokas2019b}. All bars end up with a pronounced boxy/peanut shape when viewed edge-on at the end of simulations
(Fig.~\ref{surdenxz10gyr}). Interestingly, this also applies to the highest gas fraction case ($f=0.3$) where the bar
did not buckle but still heated vertically, as discussed by \citet{Pfenniger1990} and \citet{Debattista2006}. Similar
shapes are visible in the gas component that did not convincingly buckle and it must have responded to the
changes of the mass distribution in the stars, which was always dominant in our simulations.

Our finding that models in which buckling is weak (gas fraction $f=0.2$) or even absent ($f=0.3$) can still form
boxy/peanut shapes agrees with the observational results of \citet{Erwin2017}. They find that the fraction of the
boxy/peanut bulges increases strongly with the stellar mass and in the range of stellar masses considered here it
depends rather weakly on the gas fraction, although no boxy/peanut shapes are found in galaxies with the present gas
fraction $f > 0.5$. Still, we should expect such shapes to be weaker in galaxies with higher gas fractions. The results
presented here are also consistent with the recent finding of \citet{Kruk2019} who studied the fraction of galaxies
with a boxy/peanut bulge as a function of redshift using data from the Hubble Space Telescope Cosmic Evolution
Survey and the Sloan Digital Sky Survey. They find this fraction to strongly decrease with redshift down to zero at a
redshift of $z=1$. Although no information was available concerning the gas content of higher redshift galaxies, one
can expect them to be more gas-rich than their nearby counterparts. Since higher gas content can delay the vertical
thickening of the bar, younger galaxies are expected to have lower boxy/peanut fractions.

\citet{Berentzen2007} only considered small gas fractions of 0.005-0.08 and did not see any clear trend of the bar
strength and the strength of buckling with the increasing gas content. The only difference was a marginally shorter
rise time of the bar instability in gas-richer models, which is in agreement with our result for $f=0.1$. In all of
their simulations, the stellar bars grew to the same strength and this strength then dropped by the same amount. In our
simulations (the upper panel of Fig.~\ref{a2}), this drop is smaller for higher gas fractions, thus confirming that
buckling is indeed systematically weaker. They were able to detect differences in the buckling characteristics only
when measuring the vertical asymmetry and thickening of the bar. Their bars only showed clear signatures of buckling
asymmetry for the purely collisionless case. For
the remaining gas fractions, the bars just thickened vertically retaining their vertical symmetry, which is similar
to our case of $f=0.3$. We conclude that the range of gas fractions considered by \citet{Berentzen2007} was just too
narrow to see a clear monotonic dependence of bar evolution and buckling on the gas fraction. While their gas fractions
were motivated by the values that were actually observed in disk galaxies today, it should be kept in mind that these
fractions were probably higher in the past when the bars were formed.

A larger range of gas fractions was considered in a follow-up study by \citet{Villa2010}. They were able to detect
clear trends in the dependence of the bar properties on the gas content, but they find that their results depend
strongly on the adopted gas resolution without obvious convergence. Their bars are weaker in more gas-rich disks, which
is in agreement with our findings. However, although their bar mode evolution curves have sudden drops even
for high gas fractions, their asymmetry measures surprisingly do not show any buckling signals. Also in
disagreement with our results, for higher gas fractions, but still as low as 0.08, their bars strongly decline in
strength during the later secular evolutionary phase, while their pattern speeds increase.

Our results are in broad agreement with the recent work of \citet{Gajda2018} who studied the formation and evolution of
bars in dwarf galaxies orbiting the Milky Way. Although their initial dwarf model is stable against bar formation in
isolation, the bars are tidally induced and the gas fractions considered are much higher (0.3 and 0.7), the dependence
of their properties on the gas fraction is similar. The stellar bars formed more easily and survived
longer for dwarfs containing less gas. Interestingly, the gas component preserved its axisymmetry throughout the entire
evolution.

Our conclusion that low gas fractions favor stronger buckling and the formation of pronounced boxy/peanut shapes also
agrees with the results of recent zoom-in simulations of Milky Way-like galaxies in the cosmological context from the
Auriga Project. \citet{Blazquez2019} studied 21 barred galaxies from this project and identify four cases with clear
boxy/peanut shapes and two more in the buckling phase, which all have a gas content below 0.2 at a redshift of $z=0$.
These galaxies were further studied by \citet{Fragkoudi2019}, revealing the kinematic features characteristic of
buckling bars similar to those discussed in detail in \citet{Lokas2019b}.

It remains to be investigated how the processes of star formation and feedback affect the formation of bars and
their buckling if there is a supply of gas from the halo. Although the results depend on a
number of parameters related to the subgrid physics, which are poorly constrained as of yet, one can expect that the
evolution will be somewhere in between the cases of the constant gas fractions considered here. Since the gas will be
consumed by star formation, a galaxy starting with a high gas content will increase its stellar disk mass with time and
become more susceptible to the formation of a stellar bar. Thus, such a galaxy may evolve as our $f=0.3$ case initially
and as our $f=0.1$ case in the later stages. The transition between the two regimes will depend on the time scales of
bar formation and gas consumption.


\begin{acknowledgements}
I am grateful to N. Deg for help with the GalactICS code and to the anonymous referee for useful comments.
This work was supported in part by the Polish National Science Center under grant 2013/10/A/ST9/00023.
\end{acknowledgements}

\end{document}